\documentclass[journal]{IEEEtran}
\usepackage{amsmath,amsfonts}
\usepackage{algorithm}
\usepackage{algorithmicx}
\usepackage{algpseudocode}
\usepackage{array}
\usepackage{graphicx}
\usepackage{float}
\usepackage[caption=false]{subfig}
\usepackage{textcomp}
\usepackage{stfloats}
\usepackage{url}
\usepackage{verbatim}
\usepackage{amssymb} 
\usepackage{tabularx}
\usepackage{multirow}
\usepackage{hyperref}
\usepackage{booktabs}
\usepackage{caption}
\usepackage{bm}
\def\BibTeX{{\rm B\kern-.05em{\sc i\kern-.025em b}\kern-.08em
    T\kern-.1667em\lower.7ex\hbox{E}\kern-.125emX}}
\usepackage{balance}
\usepackage{tikz}

\usepackage{color}

\long\def\comment#1{}

\def\ie{$i.e.$}
\def\eg{$e.g.$}
\def\etal{\textit{et al.} }

\newcommand{\tabincell}[2]{\begin{tabular}{@{}c#1@{}}#2\end{tabular}}

\begin{document}

\title{Towards Stealthy Backdoor Attacks against Speech Recognition via Elements of Sound}

\author{Hanbo Cai, Pengcheng Zhang, Hai Dong, Yan Xiao, Stefanos Koffas, Yiming Li
        % <-this % stops a space
%\thanks{This paper was produced by the IEEE Publication Technology Group. They are in Piscataway, NJ.}% <-this % stops a space

\thanks{Hanbo Cai and Pengcheng Zhang are with College of Computer and Information, Hohai University, Nanjing, China (e-mail: caihanbo@hhu.edu.cn, pchzhang@hhu.edu.cn).}
\thanks{Hai Dong is with School of Computing Technologies, RMIT University, Melbourne, Australia (e-mail: hai.dong@rmit.edu.au).}
\thanks{Yan Xiao is with School of Cyber Science and Technology, Sun Yat-sen University, Shenzhen, China (e-mail: xiaoyan.hhu@gmail.com).}
\thanks{Stefanos Koffas is with Delft University of Technology, Delft, Netherlands (e-mail: s.koffas@tudelft.nl).}
\thanks{Yiming Li is with Tsinghua University, Shenzhen, China and Ant Group, Hangzhou, China (e-mail: liyiming.tech@gmail.com).}
\thanks{Corresponding Author: Pengcheng Zhang (pchzhang@hhu.edu.cn).}
}

%\markboth{IEEE Transactions on Information Forensics and Security}%
%{IEEE Transactions on Information Forensics and Security}

\markboth{Preprint}%
{Preprint}


\maketitle

\begin{abstract}
Deep neural networks (DNNs) have been widely and successfully adopted and deployed in various applications of speech recognition. Recently, a few works revealed that these models are vulnerable to backdoor attacks, where the adversaries can implant malicious prediction behaviors into victim models by poisoning their training process. In this paper, we revisit poison-only backdoor attacks against speech recognition. We reveal that existing methods are not stealthy since their trigger patterns are perceptible to humans or machine detection. This limitation is mostly because their trigger patterns are simple noises or separable and distinctive clips. Motivated by these findings, we propose to exploit elements of sound ($e.g.$, pitch and timbre) to design more stealthy yet effective poison-only backdoor attacks. Specifically, we insert a short-duration high-pitched signal as the trigger and increase the pitch of remaining audio clips to `mask' it for designing stealthy pitch-based triggers. We manipulate timbre features of victim audios to design the stealthy timbre-based attack and design a voiceprint selection module to facilitate the multi-backdoor attack. Our attacks can generate more `natural' poisoned samples and therefore are more stealthy. Extensive experiments are conducted on benchmark datasets, which verify the effectiveness of our attacks under different settings ($e.g.$, all-to-one, all-to-all, clean-label, physical, and multi-backdoor settings) and their stealthiness. The code for reproducing main experiments are available at \url{https://github.com/HanboCai/BadSpeech_SoE}.
\end{abstract}


\section{Introduction}
\IEEEPARstart{S}{peech} recognition has been widely and successfully deployed in many mission-critical applications \cite{wang2022query,marras2022dictionary,hu2023exploring}. In general, obtaining well-performed speech recognition models requires training on large-scale annotated datasets and substantial hardware resources. Accordingly, developers and users usually exploit third-party resources, such as open-source datasets and checkpoints, to alleviate training burdens.

However, recent studies revealed that outsourcing (parts of) training procedures ($e.g.$, data collection) may also introduce new security risks to DNNs \cite{goldblum2022dataset}. Arguably, backdoor attack is one of the most emerging yet threatening threats \cite{backdoorsurvey}. The backdoor adversaries can implant hidden backdoors to victim DNNs by introducing a few poisoned training samples containing adversary-specified trigger patterns. The adversaries can activate the embedded backdoor via triggers during the inference process of backdoored models to maliciously manipulate their predictions. However, the backdoored models behave normally on benign testing samples. Accordingly, victim users can hardly notice backdoor threats.

Currently, most of the existing backdoor attacks are designed against image or text classification \cite{gu2019badnets,chen2022badpre,liu2022piccolo,cui2022unified,gao2023not,qi2023revisiting}. However, the backdoor analysis in speech recognition is left far behind. In particular, the few feasible attacks in this area are preliminary, whose trigger patterns are simple noises \cite{liu2018trojaning,zhai2021backdoor,shi2022audio,koffas2022can,Ye2019adversarial} or separable and distinctive audio clips \cite{qiang2022opportunistic,xin2023natural,luo2022practical}. Accordingly, these attacks are perceptible to humans or can be easily detected and alleviated by algorithms~\cite{koffas2022can,nyquist2}. It raises an intriguing question: \textit{Is it possible to design an effective attack against speech recognition that is stealthy to both human and machine detection}?

\comment{
\begin{figure*}[htbp]
\begin{minipage}[t]{0.5\linewidth}
\centering
\includegraphics[width=0.9\columnwidth]{pic/third-part.eps}
\caption{Threat scenario of the third-party training platform.}
\label{fig:third-part}
\end{minipage}%
\hfill
\begin{minipage}[t]{0.5\linewidth}
\centering
\includegraphics[width=0.9\columnwidth]{pic/auto-driver.eps}
\caption{Threat scenario of the autonomous driving.}
\label{fig:auto-driver}
\end{minipage}
\end{figure*}
}

The answer to the aforementioned question is positive. Arguably, the core of an effective and stealthy attack is to design more `natural' trigger patterns. In this paper, we generate more naturally poisoned samples by modifying the elements of sound. We tackle trigger design from two perspectives, including pitch and timbre. Specifically, we first increase the pitch of selected audio samples and then insert a short yet high-pitched signal to generate their poisoned version for the pitch-based attack. The pitch-increased background audio can hide the inserted signal due to audio masking. This method is dubbed pitch boosting and sound masking (PBSM); For the timbre-based attack, we edit the timbre features of selected samples to generate their poisoned counterparts. In particular, we design a voiceprint selection module that enables the selection of diverse timbre features for timbre transformation, to further improve its effectiveness under the multi-backdoor setting. We call this method voiceprint selection and voice conversion (VSVC). The poisoned samples generated by our PBSM and VSVC are natural and sample-specific. As such, they can bypass both human inspection and machine detection.

In conclusion, our main contributions are three-fold:

\begin{itemize}
\item We reveal the stealthiness deficiency of existing attacks against speech recognition and its potential reasons.

\item We propose two simple yet effective backdoor attacks against speech recognition ($i.e.$, PBSM and VSVC) via elements of sound. The poisoned samples of both PBSM and VSVC are more natural and therefore stealthy to both human inspection and machine detection. 

\item Extensive experiments are conducted to verify the effectiveness of our attacks under different settings ($e.g.$, all-to-one, all-to-all, clean-label, physical, and multi-backdoor settings) and their resistance to defenses. 

\end{itemize}

The rest of this paper is structured as follows. In Section~\ref{sec:relatedworks}, we briefly review related works about speech recognition and backdoor attacks. Section~\ref{sec:methods} illustrates our two stealthy backdoor attacks based on elements of sound, $i.e.$, pitch boosting and sound masking (PBSM) and voiceprint selection and voice conversion (VSVC), in details. The experimental results of our attacks are presented in Section~\ref{sec:experimenal}. We conclude this paper in Section~\ref{sec:conclusion} at the end.

\section{Related Works}
\label{sec:relatedworks}

\subsection{Speech Recognition}

Speech recognition (SR) plays a vital role in many critical applications \cite{1976speech}, allowing devices to comprehend and interpret human speech. Early speech recognition methods were mostly based on Gaussian mixture models (GMMs) and hidden Markov models (HMMs) \cite{gales2008application}. However, these methods suffered from relatively high error rates in practice.

Recently, advanced SR methods were all based on deep neural networks (DNNs) due to their high learning capacities. For example, Hinton \etal \cite{hinton2012deep} applied DNNs to acoustic modeling and achieved promising performance in the TIMIT \cite{garofolo1993timit} phoneme recognition task, marking a breakthrough in the field of speech recognition with DNNs. De \etal \cite{de2018neural} applied long short-term memory (LSTM) networks in speech recognition tasks, motivated by the strong temporal nature of speech data. Besides, inspired by the tremendous success of ResNet in image classification \cite{resnet}, Vygon \etal \cite{vygon2021learning} proposed a novel and effective keyword discovery model with the ResNet backbone. Recently, Axel \etal \cite{kwt} exploited the Transformer structure in speech recognition and achieved remarkable performance. Avi \etal \cite{EAT} proposed an end-to-end strategy without requiring pre-processing speech data to simplify the speech recognition tasks. Specifically, they adopted one-dimensional convolutional stacks and Transformer-type encoder blocks to process and classify speech data.

\subsection{Backdoor Attacks} 
Backdoor attack is an emerging yet critical training-phase threat \cite{backdoorsurvey}. In general, the adversaries intend to implant hidden backdoors into the victim model by maliciously manipulating the training procedures ($e.g.$, samples or loss). The backdoored model will behave normally on predicting benign testing samples whereas its predictions will be misled to adversary-specified target classes whenever its backdoor is activated by the trigger pattern contained in attacked testing samples.

Currently, most of the existing attacks are designed against image classification. These attacks can be divided into different sub-categories based on different criteria, as follows:

\vspace{0.3em}
\noindent \textbf{Poisoned-Label and Clean-Label Attacks.} Backdoor attacks can be divided into poisoned-label \cite{gu2019badnets,li2022untargeted,qi2023revisiting} and clean-label attacks \cite{turner2019label,souri2022sleeper,zeng2023narcissus} based on whether the target label of poisoned samples is consistent with their ground-truth one. In general, poisoned-label backdoor attacks are more effective compared to the clean-label ones since the `robust features' related to the target class contained in poisoned samples of clean-label attacks will hinder the learning of trigger patterns \cite{gao2023not}. However, clean-label attacks are more stealthy since victim users can identify and filter out poisoned training samples by examining the image-label relationship.

\vspace{0.3em}
\noindent \textbf{All-to-One and All-to-All Attacks.} We can separate existing attacks into all-to-one and all-to-all attacks based on the property of the target label \cite{gu2019badnets}. Specifically, all poisoned samples will be assigned the same target label in all-to-one attacks, while the target label of all-to-all attacks is determined based on the ground-truth one of the poisoned samples. For example, the all-to-all adversaries usually adopt $y' = (y+1) \mod K$, where $K$ is the number of all classes, $y'$ and $y$ indicate the target label and ground-truth label of the poisoned sample, respectively. Arguably, all existing (poisoned-label) backdoor attacks can be generalized to all-to-all attacks, although it will probably decrease attack effectiveness \cite{backdoorsurvey}.

\vspace{0.3em}
\noindent \textbf{Single-Backdoor and Multi-Backdoor Attacks.} Different from the single-backdoor attacks where the adversaries only implant a single backdoor to the victim models, multi-backdoor methods~\cite{gu2019badnets,xue2020one,hou2022m,salem2022dynamic} intend to embed multiple backdoors simultaneously. In general, it is non-trivial to implant multiple backdoors, although we can easily inject a single backdoor. It is mostly because the learning of one backdoor may affect that of the others \cite{salem2022dynamic}. As such, multi-backdoor attacks may fail if triggers are not `strong' enough.

\vspace{0.3em}
\noindent \textbf{Digital and Physical Attacks.} Different from previous digital attacks where all poisoned samples are obtained completely in the digital space, the physical space is also involved in their generation in the physical attacks. Chen \etal \cite{chen2017targeted} proposed the first physical backdoor attack where they exploited the glasses as physical trigger against facial recognition. A similar idea was also discussed in \cite{wenger2021backdoor}. Recently, Li \etal \cite{li2021backdoor} revealed that existing digital attacks will fail in the physical space and proposed a physical attack enhancement inspired by the expectation over transformation \cite{athalye2018synthesizing}. Most recently, Xu \etal \cite{xu2023batt} designed a more stealthy poison-only physical backdoor attack using spatial transformations ($e.g.$, rotation) with a specific parameter as trigger patterns.

\begin{figure*}[!t]
\centering
\includegraphics[width=\linewidth]{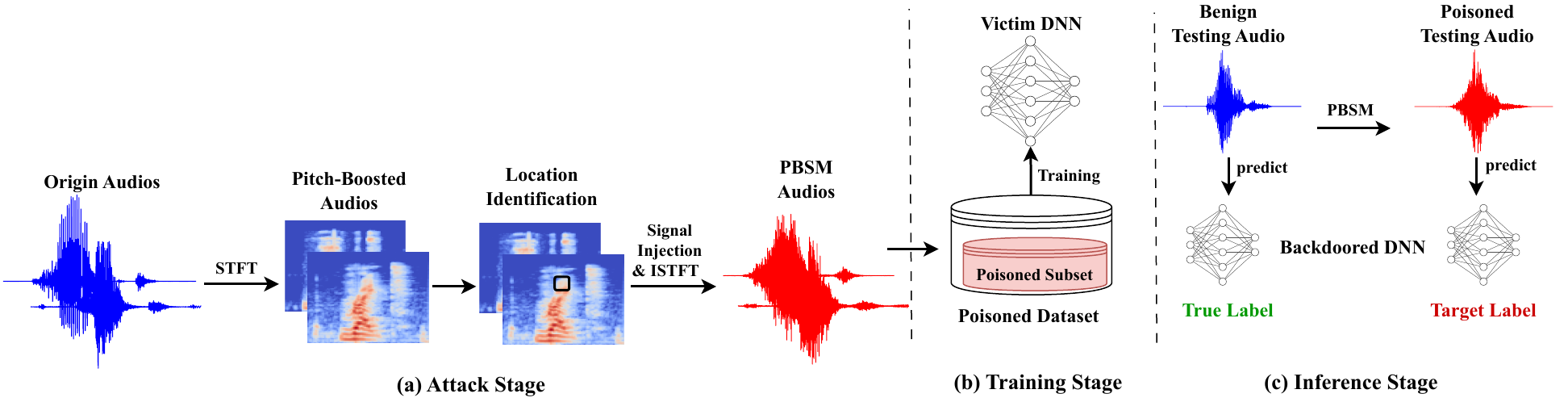}
\caption{The main pipeline of attacking via our pitch boosting and sound
masking (PBSM). The PBSM consists of three main stages, including attack, training, and inference. The attack stage is the core of PBSM, containing two steps (\ie, pitch boosting and signal injection). In the first step, we exploit short-time Fourier transform to convert the original audio from the time domain to the frequency domain and increase the pitch of the overall audio; In the second step, we identify the position of the highest-amplitude segment in the audio where we insert an adversary-specified high-pitched signal.}
\label{fig:PBSM_Pipeline}
\end{figure*}

Recently, there are also a few backdoor attacks against speech recognition. %other tasks \cite{xi2021graph,li2022few,mei2023notable}, especially 
Specifically, Liu \etal \cite{Liu2018} reversed potential training samples of a given speech recognition model, based on which to implant hidden backdoors; Ye \etal \cite{Ye2019adversarial} designed trigger patterns based on audio steganography; Zhai \textit{et al.}~\cite{zhai2021backdoor} designed the first backdoor attack against speaker verification via clustering techniques; Koffas \etal exploited ultrasonic pulses as audio triggers; In \cite{qiang2022opportunistic,luo2022practical,xin2023natural}, sounds from the natural environment ($e.g.$, music and noises) were adopted as trigger patterns; Shi \etal \cite{shi2022audio} developed an optimization scheme to generate more effective audio triggers; Most recently, a concurrent work \cite{koffas2023going} designed stealthy style-based triggers for audio backdoor attacks via style transformations. However, all existing attacks are perceptible to humans or can be easily detected and alleviated by algorithms. How to design an effective backdoor attack against speech recognition that is stealthy to both human and machine detection is still an important open question and worth further exploration.

\section{The Proposed Methods}
\label{sec:methods}
The sound elements primarily include pitch, timbre, and loudness~\cite{erickson1975sound}. In this paper, we discuss how to design more natural yet effective acoustic trigger patterns based on pitch and timbre, respectively. We omit the loudness-type trigger design since it has minor variation and therefore may not contain sufficient information for effective backdoor attacks.

\subsection{Preliminaries}
\vspace{0.3em}
\noindent \textbf{Threat Model.} In this paper, we focus on \emph{poison-only} backdoor attacks against speech recognition, where the adversaries can only modify their released poisoned training dataset. The victim users will exploit the poisoned dataset to train their models with user-specified settings. Accordingly, we assume that the adversaries cannot change and have no information on the training process (\eg, model structure, loss, and training schedule). This is one of the most difficult settings for backdoor attacks, with the most expansive threat scenarios (\eg, using third-party samples, training facilities, or models) \cite{backdoorsurvey}.

\vspace{0.4em}
\noindent \textbf{Adversary's Goals.} In summary, the backdoor adversaries have three main goals, including \textbf{(1)} effectiveness, \textbf{(2)} stealthiness, and \textbf{(3)} persistence. Specifically, effectiveness requires that backdoored models can predict poisoned testing samples as the adversary-specified target label, no matter what their ground-truth label is; Stealthiness ensures that the attack cannot be detected by human inspection or simple machine detection. For example, trigger patterns should be stealthy and the poisoning rate should be small; Persistence seeks that the attack is still effective under more difficult settings (\eg, under potential adaptive defenses and physical-world settings).

\vspace{0.4em}
\noindent \textbf{The Main Pipeline of Poison-Only Backdoor Attacks.} In general, how to generate the poisoned dataset $\hat{\mathcal{D}}$ given its benign version $\mathcal{D} = \{ (\bm{x}_i,y_i) \}_{i=1}^{N}$ is the main problem of poison-only backdoor attacks. Considering a classification problem with $K$-categories, the $\hat{\mathcal{D}}$ contains two separate subsets, including the benign subset $\mathcal{D}_b$ and the poisoned subset $\mathcal{D}_p$ (\ie, $\hat{\mathcal{D}} = \mathcal{D}_b \cup \mathcal{D}_p$). Specifically, $\mathcal{D}_b$ is randomly sampled from $\mathcal{D}$ containing $(1-\gamma) \cdot N$ samples, where $\gamma$ is dubbed `poisoning rate'. $\mathcal{D}_p \triangleq \left\{ \left(G_x(\bm{x}), G_y(y) \right)| (\bm{x}, y) \in \mathcal{D} \backslash \mathcal{D}_b \right\}$, where $G_x: \mathcal{X} \rightarrow \mathcal{X}$ and $G_y: \mathcal{Y} \rightarrow \mathcal{Y}$ are adversary-assigned poisoned instance generator and poisoned label generator, respectively. For example, $G_x(\bm{x}) = \bm{x} + \bm{t}$ where $\bm{t}$ is the trigger based on additive noises \cite{li2021invisible}; $G_y(y) = y_T$ where $y_T$ is the target label in all-to-one attacks \cite{backdoorsurvey}, $G_y(y) = (y+1) \mod K$ in most of the existing all-to-all attacks \cite{gu2019badnets}. After $\hat{\mathcal{D}}$ is generated and released, the victim users will use it to train their model $f_{\bm{\theta}}: \mathcal{Y} \rightarrow [0,1]^{K}$ via $\min_{\bm{\theta}} \sum_{(\bm{x},y) \in \hat{\mathcal{D}}} \mathcal{L}(f_{\bm{\theta}}(\bm{x}), y)$.

\subsection{Attack via Pitch Boosting and Sound Masking}
\label{sec:PBSM}

Arguably, the most straightforward approach to designing pitch-type triggers is to insert sound clips with a very high (or low) frequency in a random position of the victim audio. However, these triggers can be easily filtered out by removing clips with the highest and lowest frequencies. Besides, these triggers are also perceptible to humans since the inserted trigger is most likely different from its surrounding audio clips in the poisoned samples. To tackle these problems, in this paper, we propose to first increase the pitch of selected audio samples and then insert a short yet high-pitched signal to the position with the highest sound energy. This method is dubbed attack via pitch boosting and sound masking (PBSM).

The pitch boosting makes our attack resistant to trigger filtering (as shown in our experiments). The filtering cannot decrease the pitch of poisoned audio since these triggers are natural, although it may remove the high-pitched short signal. Besides, our insertion strategy improves the stealthiness of triggers for both human inspection and machine detection. Specifically, the inserted high-pitched signal is less perceptible to humans due to sound masking while it can bypass classical detection methods based on finding common audio clips since the insert position is usually sample-specific. In other words, different poisoned samples have different insert positions.

In general, our PBSM has two main steps, including \textbf{(1)} pitch boosting and \textbf{(2)} signal injection, to generate poisoned samples. The details of this process is described in Algorithm \ref{alg:pbsm} and the main pipeline of PBSM is shown in Figure \ref{fig:PBSM_Pipeline}.

\vspace{0.3em}
\noindent \textbf{Step 1: Pitch Boosting.} A feasible method for pitch boosting is to increase the frequency of selected audio samples. Accordingly, we first perform a short-time Fourier transform (STFT) \cite{orfanidis1995introduction} on the original audio to convert it from the time domain to the frequency domain. After that, in the frequency domain, we multiply the original frequency values by an adversary-specified pitch-shifting coefficient $p$ ($p>1$), leading to a new audio waveform with a boosted pitch. Specifically, we can express the short-time Fourier transform as $\bm{x_f}= \mathcal{F}(\bm{x})$ (Line 1 in Algorithm \ref{alg:pbsm}), where $\bm{x_f}$ is the frequency-domain representation of $\bm{x}$. The process of increasing pitch can be expressed as  $\bm{x_P}= p \cdot \sum_{i=0}^{L_p}\bm{x_f}^{(i)}$ (Line 3 in Algorithm \ref{alg:pbsm}). Specifically, in the aforementioned equation, $L_p$ represents the number of points in the frequency domain, the transformation factor $p$ is represented as $p = 2^{n\_p/12}$, and $n\_p$ denotes the number of semitones (\ie, the step of pitch shifting).

\vspace{0.3em}
\noindent \textbf{Step 2: Signal Injection.} This process consists of two main stages, including \textbf{(1)} location identification and \textbf{(2)} signal insertion. In the first stage, we identify the location of the high-amplitude segments in the audio signal. We select the high-amplitude clips since they have stronger energy and can provide better masking effects. Specifically, to find these positions, we iterate through each audio segment to identify the position of the segment with the highest energy in the entire audio sample. The position $T$ of high-amplitude segments can be obtained by: $T=\underset{i}{\operatorname{argmax}}( \sum_{i}^{i+L}|\bm{x_P^{(i)}}|) + L$ (Line 5 in Algorithm \ref{alg:pbsm}), where $L$ is the high-amplitude length. In the second stage, we insert an adversary-specified high-pitched signal $\bm{h}$ in the selected position $T$. Specifically, this process can be denoted by $\bm{x_r} = \bm{x_P}^{(T)} \oplus \bm{h}$ (Line 6 in Algorithm \ref{alg:pbsm}), where $\bm{x_r}$ is the inserted audio signal after signal injecting, $\bm{x_P}^{(T)}$ is the audio segment at position $T$, and $\oplus$ denotes the injection operation with the high-pitched signal $\bm{h}$. We conduct the inverse Fourier transformation $\mathcal{F}^{-1}$ \cite{orfanidis1995introduction} to obtain poisoned audio with pitch-type triggers by turning frequency-domain signals back to the time domain (Line 7 in Algorithm \ref{alg:pbsm}).

\begin{algorithm}[!t]
  \caption{The algorithm of pitch boosting and sound masking (PBSM).}
  \label{alg:pbsm}
  \begin{algorithmic}[1]
    \Require Benign audio $\bm{x}$ and high-pitch signal $\bm{h}$.
    \State  $\bm{x_f}= \mathcal{F}(\bm{x})$ // Short-time Fourier transformation.
    \For{$i$ in range($\bm{x_f}$)}
    \State   $\bm{x_P}= p \cdot \sum_{i=0}^{{L_p}}\bm{x_f}^{(i)}$ // Pitch boosting in the frequency domain.
    \EndFor
    \State    $T=\underset{i}{\operatorname{argmax}}( \sum_{i}^{i+L}|\bm{x_P}^{(i)}|) + L$ // Calculate the position of high-amplitude segments.   
    \State $\bm{x_r} = \bm{x_P}^{(T)} \oplus \bm{h}$ // Insert a high-pitch signal.
    %\Statex \textbf{\textit{//Inverse Fourier transform back to the time domain}}
    \State $\bm{x_t}= \mathcal{F}^{-1}(\bm{x_r})$ // Inverse Fourier transformation.
    \Ensure Poisoned audio $\bm{x_t}$.
    %\Statex \textbf{Return:} Poisoned audio $\bm{x_t}$.
  \end{algorithmic}
\end{algorithm}

\subsection{Attack via Voiceprint Selection and Voice Conversion}

To design timbre-type triggers, we can exploit a `timbre transformer' trained on the audios of an adversary-specified target people (\eg, the adversary himself) for voice conversion \cite{voiceconversion}. Specifically, we can assign the poisoned instance generator $G$ as the (pre-trained) timbre transformer. 

Assume that there are multiple timbre candidates for selection. Arguably, the design of timbre-type single-backdoor attacks is straightforward, where the adversaries can arbitrarily choose any single timbre they desire. However, the design of multi-backdoor attacks is challenging since simply selecting multiple timbres at random to design triggers has limited attack effectiveness (as we will show in the experiments). It is mostly because there can be many similarities between timbres. On the one hand, this similarity makes it harder for DNNs to learn backdoors, since similar poisoned samples have different (target) labels. On the other hand, this similarity may lead to false backdoor activation by attacked models at the inference process. Motivated by these understandings, we propose a \emph{voiceprint selection module} to alleviate these challenges.

\begin{figure*}[htbp]
\centering
\includegraphics[width=1.01\linewidth]{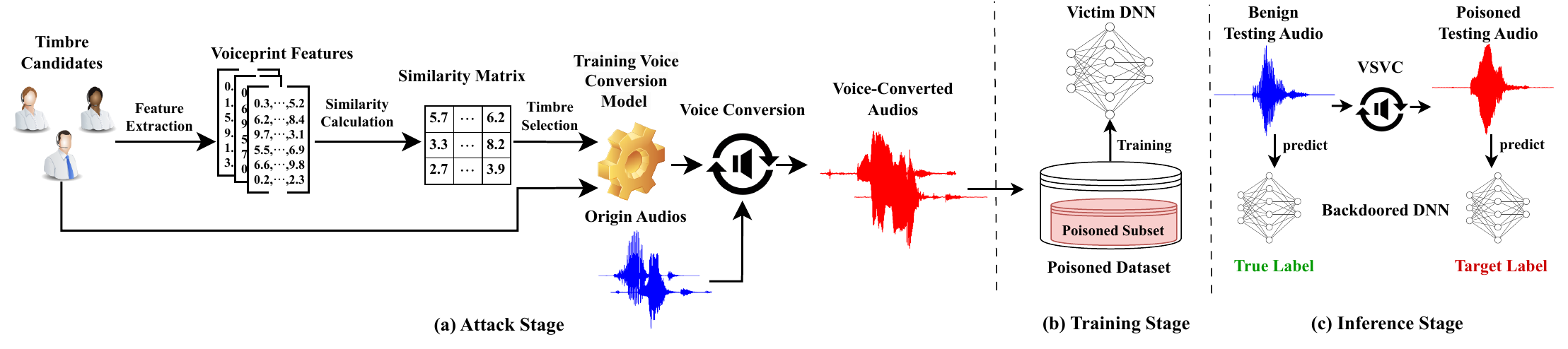}
\caption{The main pipeline of attacking via our voiceprint selection and voice conversion (VSVC). The VSVC consists of three main stages, including attack, training, and inference. The attack stage is the core of VSVC, containing four steps (\ie, feature extraction, similarity calculation, timbre selection, and voice conversion). In the first step, we adopt X-vectors to extract voiceprint features of each timbre candidate; In the second step, we measure the similarity of each timbre pair based on their distance; In the third step, we select the desired number of timbres based on the principle of smallest similarity; In the fourth step, we generate the poisoned training dataset of the (multi-backdoor) timbre-type attack via voice conversion.}
\label{fig:VSVC_Pipeline}
\end{figure*}

In general, our voiceprint selection module consists of three main stages, including \textbf{(1)} feature extraction, \textbf{(2)} similarity calculation, and \textbf{(3)} timbre selection. The main pipeline of our voiceprint selection and voice conversion (VSVC) is shown in Figure \ref{fig:VSVC_Pipeline}. Its technical details are as follows. 

\vspace{0.3em}
\noindent \textbf{Step 1: Feature Extraction.} Following the most classical method, we exploit X-vectors~\cite{xvector} to extract voiceprint features of each timbre candidates, \ie, $\bm{S_e^{(k)}} \gets V(C_k)$, where $C_k$ is the speech data for the $k$-th speaker, $V$ denotes the process of extracting X-vectors converting each speech into a $d$-dimensional feature vector, and $\bm{S_e^{(k)}}$ represents the voiceprint embedding for the $k$-th speaker. For $K$ candidates, we ultimately obtain a matrix $\bm{S_{e}} = [\bm{S_{e}^{(1)}}, ..., \bm{S_{e}^{(K)}}] \in \mathbb{R}^{d \times K}$ with $d$ rows and $K$ columns (Lines 1-3 in Algorithm \ref{alg:vsvc}) .

\vspace{0.3em}
\noindent \textbf{Step 2: Similarity
Calculation.} In this step, we calculate the distance between the features of each timbre pair $(i, j)$ as their similarity. Specifically, to represent the voiceprint distances between $K$ candidates, we construct a similarity matrix $Sim$ of size $K^2$, where each element $Sim[i][j]$ is computed as $d\left(\bm{S_{e}^{(i)}}, \bm{S_{e}^{(j)}}\right)$ (Lines 4-7 in Algorithm \ref{alg:vsvc}) with the distance metric $d$. In this paper, we assign $d$ as $\ell_2$-norm for simplicity.

\vspace{0.3em}
\noindent \textbf{Step 3: Timbre Selection.} In this step, we select $M$ candidates with maximum distances, based on the similarity matrix calculated in the previous step. We design a greedy search method to select suitable candidates (Lines 9 in Algorithm \ref{alg:vsvc}). Specifically, we select the two timbres with the greatest distance in the similarity matrix to add to the selected set $\mathcal{C}_M$. After that, we select the timbre that has the greatest distance from all the timbres in the selected set from the remaining candidates and add it to the selected set. We repeat the above process until the selected set $\mathcal{C}_M$ contains $M$ timbres.

\vspace{0.3em}
\noindent \textbf{Step 4: Generating the Poisoned Dataset via Voice Conversion.} In this step, we first train a voice conversion model $G$ (Line 10 in Algorithm \ref{alg:vsvc}), based on the the selected set $\mathcal{C}_M$ obtained in the previous step. For each audio $\bm{x}$, $G(\bm{x}, i)$ can convert its timbre to that of $i$-th element in $\mathcal{C}_M$. After that, we select $M$ adversary-specified target labels $\{y_T^{(i)}\}_{i=1}^M$. Each target label is associated with a timbre backdoor. The generated poisoned dataset $\hat{\mathcal{D}}$ contains $(M+1)$ disjoint subsets, including one benign subset $\mathcal{D}_b$ and $M$ poisoned subsets (\ie, $\{\mathcal{D}_p^{(i)}\}_{i=1}^M$). Specifically, $\mathcal{D}_p^{(i)} \triangleq \{(G(\bm{x}, i), y_T^{(i)})| (\bm{x}, y) \in \mathcal{D}_s^{(i)}\}$ where $\mathcal{D}_s^{(i)} \subset \mathcal{D}$, $\mathcal{D}_s^{(i)} \cap \mathcal{D}_s^{(j)} = \emptyset \ (\forall i \neq j)$ (Lines 11-14 in Algorithm \ref{alg:vsvc}), and $\mathcal{D}_b = \mathcal{D} - \bigcup_{i=1}^M \mathcal{D}_s^{(i)}$ (Line 15 in Algorithm \ref{alg:vsvc}). In particular, $\gamma_i \triangleq \frac{|\mathcal{D}_s^{(i)}|}{|\mathcal{D}|}$ is dubbed as the poisoning rate of $i$-th timbre-type backdoor.

\begin{algorithm}[!t]
\caption{The algorithm of voiceprint selection and voice conversion (VSVC).}
\label{alg:vsvc}
  \begin{algorithmic}[1]
    \Require Benign dataset $\mathcal{D}$, the number of backdoors $M$, poisoning rates for $M$ poisoned subsets $\{\gamma_i\}_{i=1}^M$, the number of timbre candidates $K$, timbre candidates $\mathcal{C}_K$, and target labels $\left\{y_T^{(i)}\right\}_{i=1}^M$.
    \For{$k$ in range($K$)}
    \State   $S\_e$ $\gets$ \textit{V}($\mathcal{C}_k$); // Extract the voiceprint features.
    \EndFor
    \For{$i$ in range($K$)}
        \For{$j$ in range($K$)}
        \State $Sim[i][j] = d(\bm{S\_e^{(i)}},\bm{S\_e^{(j)}})$ // Generate a similarity matrix based on distance between each pair.
        \EndFor
    \EndFor
    \State $\mathcal{C}_M  = GreedySearch(Sim, M)$ // Select $M$ suitable timbre candidates.
    \State $G$ = Train($\mathcal{C}_M$) // Training the voice conversion model.
    \For{$i$ in range($M$)}
    \State $\mathcal{D}_s^{(i)} \triangleq  Extract(\mathcal{D}-\bigcup_{j=1}^{i-1} \mathcal{D}_s^{(j)}, \gamma_i \cdot |\mathcal{D}|)$  // Extract $M$ disjoint subsets for poisoning.
    \State $\mathcal{D}_p^{(i)} \triangleq \{(G(\bm{x}, i), y_T^{(i)})| (\bm{x}, y) \in \mathcal{D}_s^{(i)}\}$ //Generate voice-converted poison samples.
    \EndFor
    \State $\mathcal{D}_b = \mathcal{D} - \bigcup_{i=1}^M \mathcal{D}_s^{(i)}$
    \State $\hat{\mathcal{D}} = \mathcal{D}_b \cup \left(\bigcup_{i=1}^M \mathcal{D}_p^{(i)}\right)$
    \Ensure The poisoned dataset $\hat{\mathcal{D}}$ generated by our VSVC.
  \end{algorithmic}
\end{algorithm}

\section{Experiments}
\label{sec:experimenal}
\subsection{Main Settings}
\label{experimentalsetting}
%\vspace{0.3em}
\noindent \textbf{Dataset Description.} We adopt the most classical benchmark, \ie, Google Speech Command dataset~\cite{speechcmd}, for our experiments. It consists of 30 common English speech commands. Each command is spoken by multiple individuals in various ways, resulting in a total of 64,728 samples. The dataset has a 16kHz sampling rate where each sample lasts approximately one second. Specifically, we selected 23,726 audios with 10 labels (dubbed `SPC-10') and 64,721 audios with 30 labels (dubbed `SPC-30') for a comprehensive comparison.

\vspace{0.3em}
\noindent \textbf{Baseline Selection.} We compared our PBSM and VSVC with four representative speech backdoor attacks, including \textbf{(1)} position-independent backdoor attack (PIBA)~\cite{shi2022audio}, \textbf{(2)} dual adaptive backdoor attack (DABA)~\cite{qiang2022opportunistic}, \textbf{(3)} backdoor attack with ultrasonic (dubbed `Ultrasonic') \cite{koffas2022can},  and \textbf{(4)} backdoor attack via style transformation (dubbed `JingleBack') \cite{koffas2023going}.

\vspace{0.3em}
\noindent \textbf{Model Structures.} As the poison-only backdoor attacks, we assume that the adversaries have no information about the victim model. To evaluate the effectiveness across different DNNs, we evaluate all attacks under four classical and advanced DNN structures, including LSTM \cite{LSTM}, ResNet-18 \cite{resnet}, KWT \cite{kwt}, and EAT \cite{EAT}. Specifically, LSTM and ResNet-18 are classical models designed for sequential and non-sequential data, respectively; KWT and EAT are advanced speech recognition models, where KWT exploited transformer structure and EAT was designed in an end-to-end manner.

\begin{table*}[!t]
\centering
\caption{The benign accuracy (BA) and attack success rate (ASR) of methods on the SPC-10 dataset.}
\vspace{-0.5em}
\label{tab:SPC-10-main}
\scalebox{1.2}{
\begin{tabular}{c|c|c|cc|cc|cc}
\toprule
Model$\downarrow$         & Metric$\downarrow$,  Method$\rightarrow$ & No Attacks  & PIBA  & DABA & Ultrasonic & JingleBack & \tabincell{}{PBSM \\ (Ours)}  & \tabincell{}{VSVC \\ (Ours)}  \\ \hline \rule{0pt}{8pt}
\multirow{2}{*}{LSTM}     & BA (\%)                                   & 93.68      & 93.54 & 92.13 & 93.21 & 92.63 & 93.32 & 93.43 \\
                          & ASR (\%)                                  & --         & 95.23 & 99.76 & 98.61 & 91.31 & 92.11 & 99.61 \\ \hline \rule{0pt}{8pt}
\multirow{2}{*}{ResNet-18} & BA (\%)                                   & 95.11      & 94.32 & 94.10 & 94.97 & 94.55 & 94.85 & 94.93 \\
                          & ASR (\%)                                  & --          & 96.43 & 99.87 & 99.33 & 95.52 & 95.78 & 97.57 \\ \hline \rule{0pt}{8pt}
\multirow{2}{*}{KWT}      & BA (\%)                                   & 91.35       & 90.21 & 90.10 & 91.11 & 91.19 & 91.27 & 90.96 \\
                          & ASR (\%)                                  & --          & 96.24 & 99.54 & 97.13 & 91.52 & 94.39 & 99.22 \\ \hline \rule{0pt}{8pt}
\multirow{2}{*}{EAT}      & BA (\%)                                   & 93.33       & 93.21 & 92.61 & 93.12 & 93.10 & 93.23 & 93.31 \\
                          & ASR (\%)                                  & --          & 97.32 & 99.21 & 99.12 & 87.39 & 90.13 & 92.32 \\
\bottomrule
\end{tabular}}
\vspace{-0.3em}
\end{table*}

% \usepackage{multirow}
% \usepackage{multirow}
\begin{table*}[!t]
\centering
\caption{The benign accuracy (BA) and attack success rate (ASR) of methods on the SPC-30 dataset.}
\vspace{-0.5em}
\label{tab:SPC-30-main}
\scalebox{1.2}{
\begin{tabular}{c|c|c|cc|cc|cc}
\toprule
Model$\downarrow$         & Metric$\downarrow$, Method$\rightarrow$ & No Attacks  & PIBA  & DABA & Ultrasonic & JingleBack & \tabincell{}{PBSM \\ (Ours)}  & \tabincell{}{VSVC \\ (Ours)} \\ \hline \rule{0pt}{8pt}
\multirow{2}{*}{LSTM}     & BA (\%)                                  & 92.62            & 92.51 & 91.18 & 92.13 & 92.57 & 92.56 & 91.91 \\
                          & ASR (\%)                                 & --               & 95.04 & 99.16 & 98.12 & 98.45 & 96.21 & 98.01 \\ \hline \rule{0pt}{8pt}
\multirow{2}{*}{ResNet-18} & BA (\%)                                  & 95.20           & 93.21 & 92.13 & 94.32 & 94.76 & 94.71 & 94.85 \\
                          & ASR (\%)                                 & --               & 98.34 & 99.98 & 97.53 & 93.39 & 96.63 & 93.01 \\ \hline \rule{0pt}{8pt}
\multirow{2}{*}{KWT}      & BA (\%)                                  & 91.13            & 90.62 & 89.19 & 90.33 & 90.20 & 90.45 & 90.21 \\
                          & ASR (\%)                                 & --               & 94.21 & 99.45 & 97.13 & 93.54 & 94.02 & 97.03 \\ \hline \rule{0pt}{8pt}
\multirow{2}{*}{EAT}      & BA (\%)                                  & 94.51            & 94.33 & 93.13 & 94.23 & 94.35 & 94.01 & 94.38 \\
                          & ASR (\%)                                 & --               & 92.12 & 99.43 & 95.32 & 81.06 & 92.51 & 93.12 \\
\bottomrule
\end{tabular}}
\vspace{-0.5em}
\end{table*}

\vspace{0.3em}
\noindent \textbf{Attack Setup.} For all attacks, we set the poisoning rate to 1\% and randomly select the `left' as the target label. For our PBSM method, we increase the pitch by 5 semitones. The length of high-amplitude segments is set to 100 milliseconds. For our VSVC method, we select the VCTK dataset \cite{VCTK} as the timbre candidates dataset and we employ StarGANv2-VC~\cite{li2021starganv2} as the voice conversion framework. In particular, we evaluate the single-backdoor VSVC in our main experiments for a fair comparison. The results of multi-backdoor VSVC are included in Section \ref{sec:mutitarget-attack}; For DABA~\cite{qiang2022opportunistic} and PIBA~\cite{shi2022audio}, we follow the same settings described in their original papers; For the ultrasonic attack \cite{koffas2022can}, we set the duration of the trigger to 100 milliseconds; For JingleBack ~\cite{koffas2023going}, we exploit the third style used in its paper since it led to the best attack performance. Note that this method may reach better stealthiness if we use other styles introduced in their paper, whereas it will decrease its attack effectiveness as the sacrifice.

\vspace{0.3em}
\noindent \textbf{Training Setup.} We extract the log-Mel spectrogram of each audio sample as an input feature, which can graphically characterize a person’s speech feature in a combination of temporal and frequency dimensions. All models are trained for 100 epochs. We set the learning rate of EAT and LSTM as 0.0001 and 0.005, respectively. We set the learning rate of the remaining models as 0.01. As for the optimizer selection, the EAT and KWT models are trained using the Adam optimizer, while the default optimizer for the other models is SGD. We run each experiment three times and calculate their average to reduce the side-effects of randomness.

\vspace{0.3em}
\noindent \textbf{Training Facilities.} We conduct all our experiments on a server running Ubuntu 18.04, equipped with a single NVIDIA GeForce RTX 3090 GPU with 24GB of VRAM.

\vspace{0.3em}
\noindent \textbf{Evaluation Metrics.} Following the most classical settings in existing works \cite{backdoorsurvey}, we adopt benign accuracy (BA) and attack success rate (ASR) to evaluate the effectiveness of all attacks. Specifically, the BA measures the proportion of benign testing samples that can be correctly classified, while the ASR denotes the proportion of poisoned testing samples that can be maliciously predicted as the target label. The higher the BA and the ASR, the more effective the attack; To evaluate the stealthiness, we invite 10 people to identify whether the poisoned audios (5 for each attack) of an attack sounded natural. The proportion of poisoned samples that are regarded as natural audios by humans is dubbed natural rate (NC). The higher the NC, the more stealthy the attack.

\subsection{Main Results}
\label{sec:mainresults}

\noindent \textbf{Attack Effectiveness.} As shown in Table \ref{tab:SPC-10-main}-\ref{tab:SPC-30-main}, the attack success rates of our PBSM and VSVC are sufficiently high ($>90\%$) in all cases on both SPC-10 and SPC-30 datasets. The attack performance of our VSVC is on par with or even better than all baseline attacks except for DABA. For example, the ASR of VSVC is 8\% higher than that of JingleBack in attacking LSTM and KWT on the SPC-10 dataset. Besides, our attacks have minor adverse effects on benign accuracy. The decreases of benign accuracy compared to the model training with benign dataset are less than 1\% in all cases for our attacks. In contrast, both DABA and JingleBack have a relatively high impact on benign accuracy. These results verify the effectiveness of our attacks.

\begin{table}[!t]
\centering
\caption{The natural rates (\%) calculated by human validation of samples generated by different methods. }
\vspace{-0.5em}
\label{tab:human-validation}
\scalebox{0.95}{
\begin{tabular}{c|cccc|cc}
\toprule
Benign & PIBA & DABA & Ultrasonic & JingleBack & \tabincell{}{PBSM \\ (Ours)} & \tabincell{}{VSVC \\ (Ours)}  \\ \hline \rule{0pt}{8pt}
100  & 0  & 0  & 100      & 0        & 98 & 100 \\
\bottomrule
\end{tabular}}
\end{table}

\begin{figure*}[htbp]
\centering
\subfloat[left\_Benign]{\label{fig:subfig:left}
\includegraphics[width=0.25\linewidth]{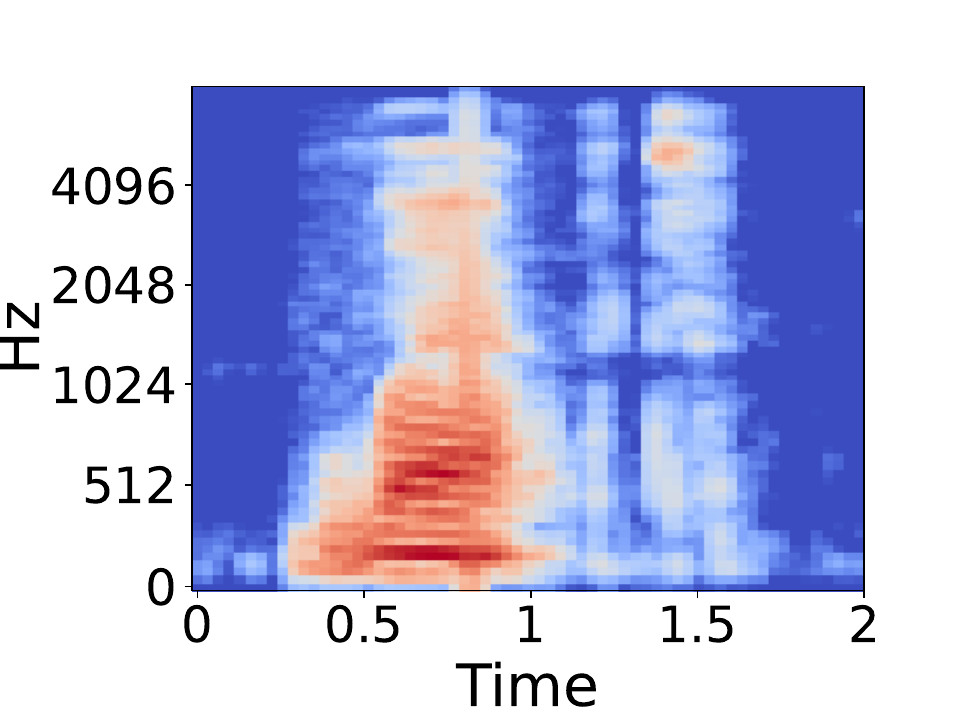}}
\subfloat[left\_PIBA]{\label{fig:subfig:left_PIBA}
\includegraphics[width=0.25\linewidth]{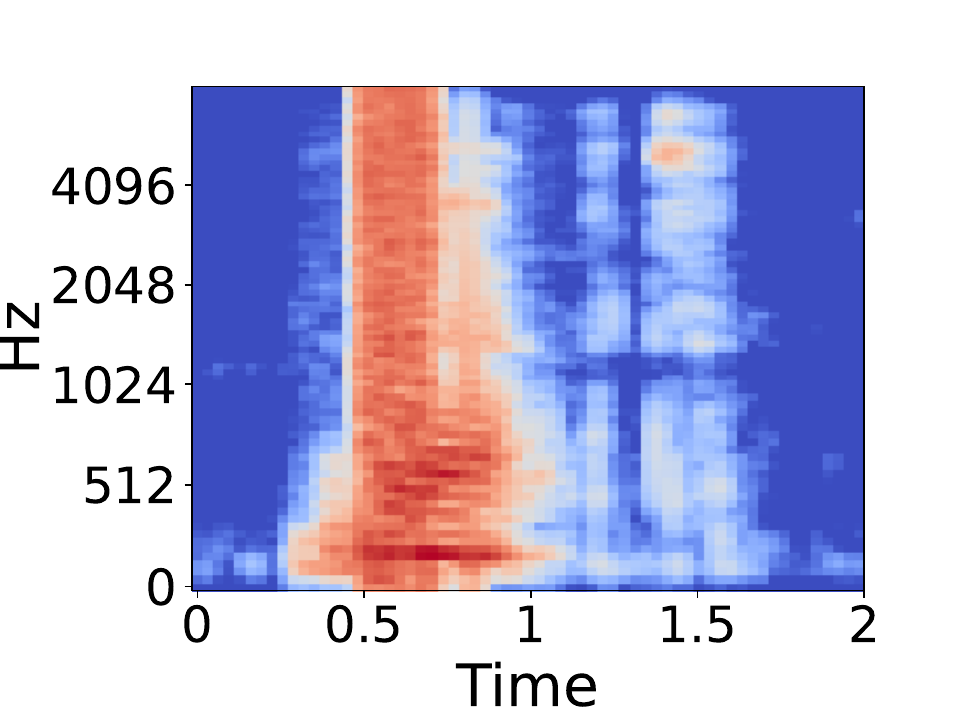}}
\subfloat[left\_DABA]{\label{fig:subfig:left_DABA}
\includegraphics[width=0.25\linewidth]{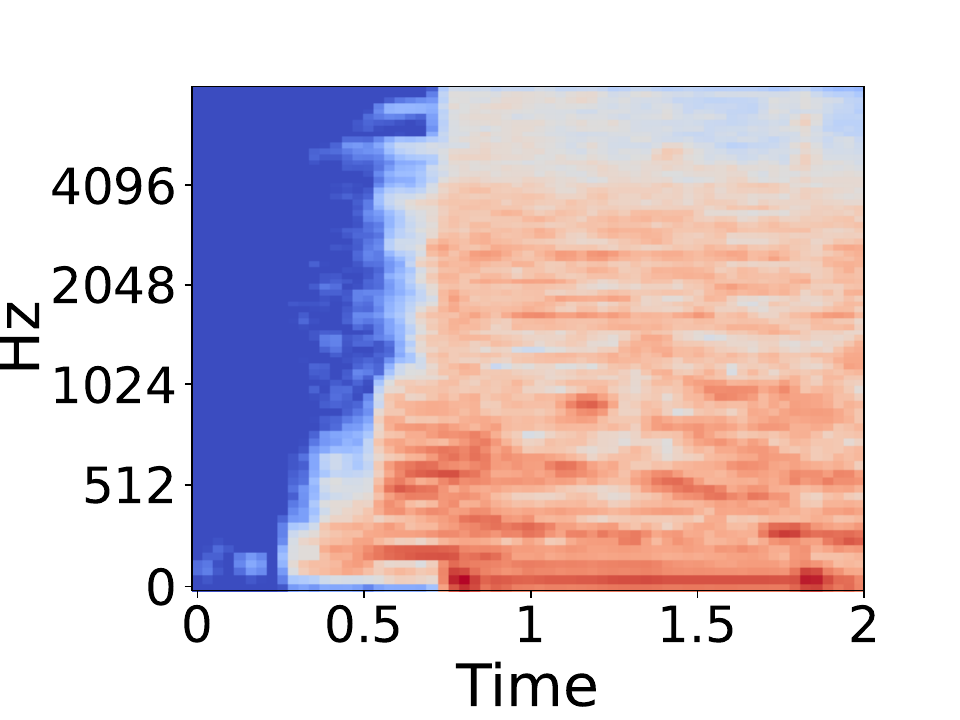}}
\subfloat[left\_Utralsonic]{\label{fig:subfig:left_Utral}
\includegraphics[width=0.25\linewidth]{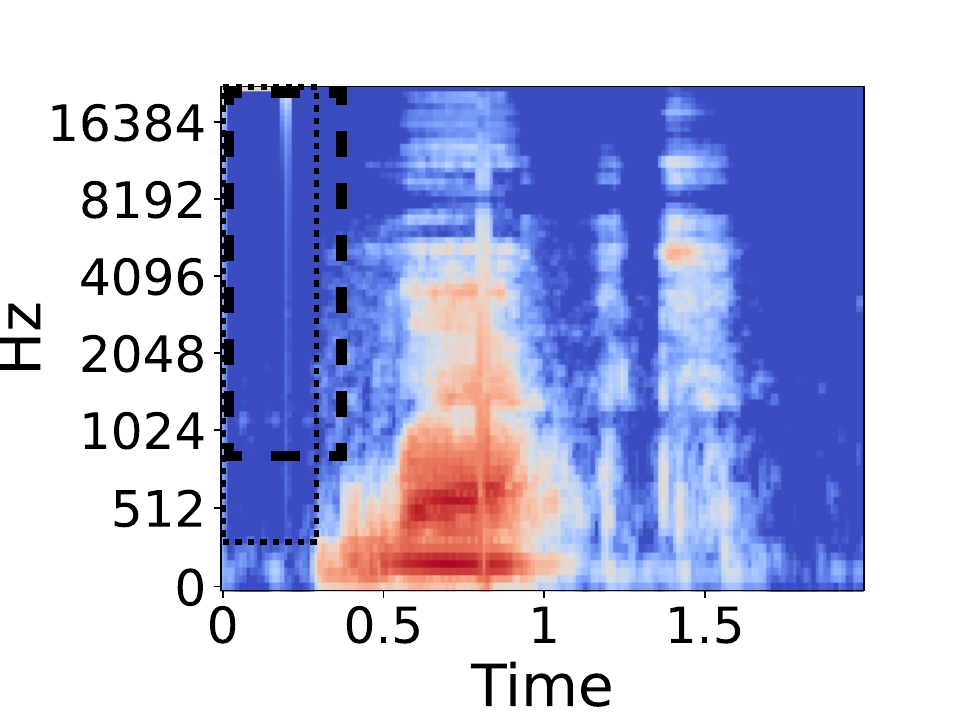}}
\vspace{-3.9mm}
\hfill
\subfloat[left\_JingleBack]{\label{fig:subfig:left_Style}
\includegraphics[width=0.25\linewidth]{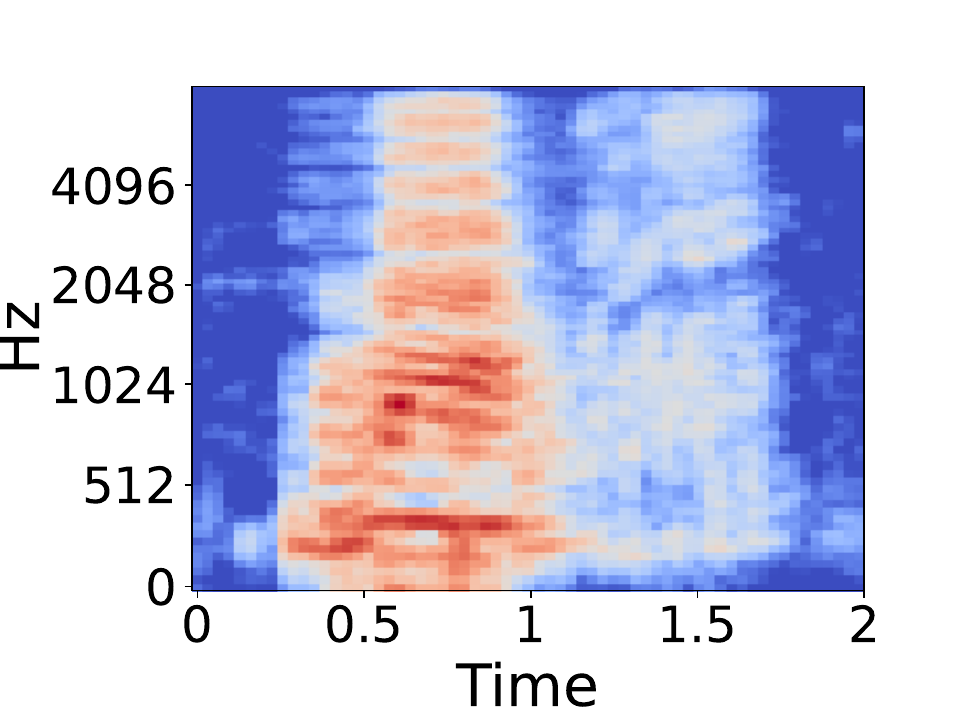}}
\subfloat[left\_Pitch]{\label{fig:subfig:left_Pitch}
\includegraphics[width=0.25\linewidth]{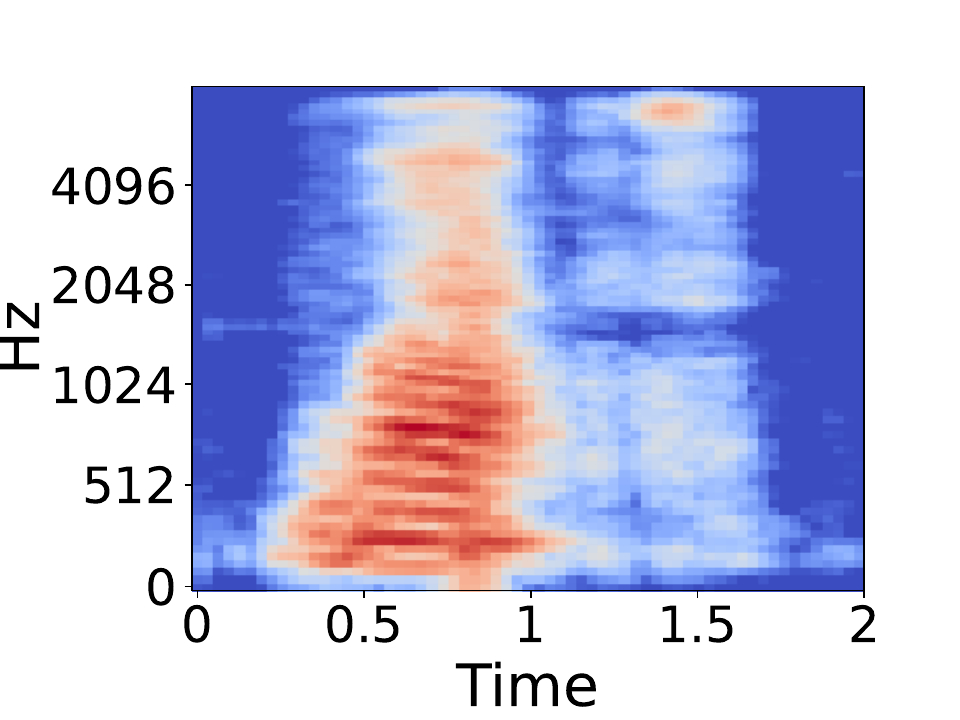}}
\subfloat[left\_PBSM]{\label{fig:subfig:left_PBSM}
\includegraphics[width=0.25\linewidth]{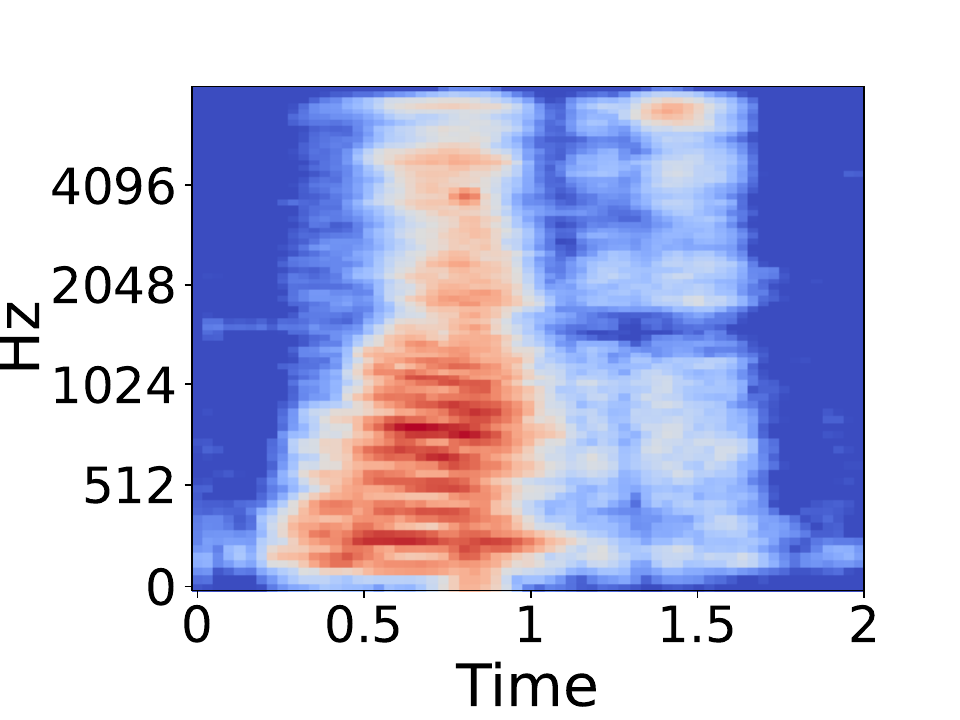}}
\subfloat[left\_VSVC]{\label{fig:subfig:left_VSVC}
\includegraphics[width=0.25\linewidth]{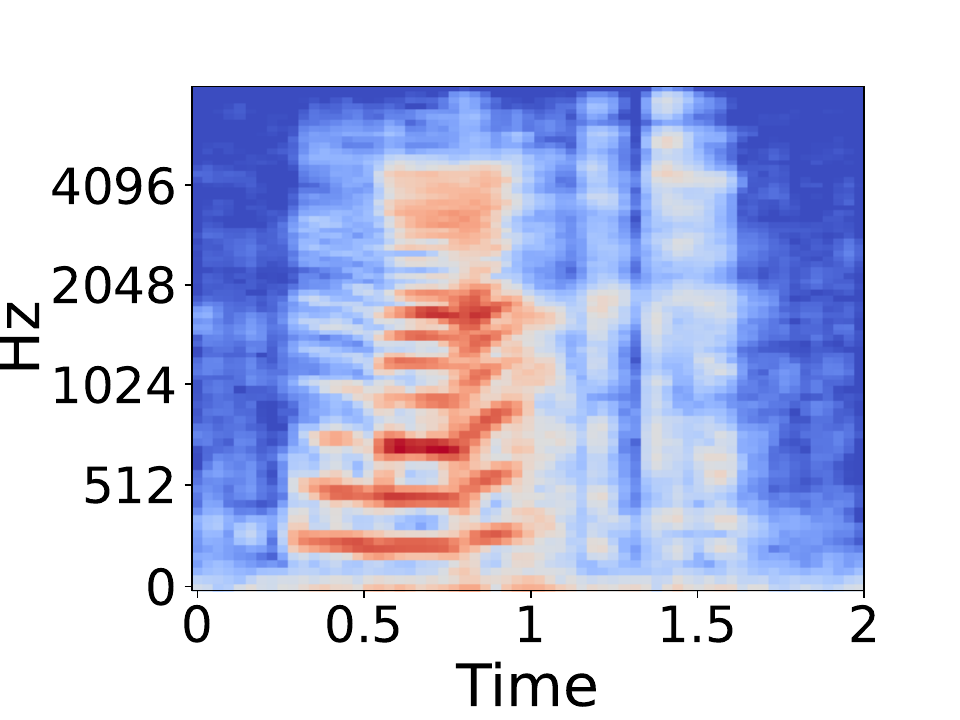}}
\vspace{-3.9mm}
\hfill
\subfloat[right\_Benign]{\label{fig:subfig:right}
\includegraphics[width=0.25\linewidth]{pic/left.pdf}}
\subfloat[right\_PIBA]{\label{fig:subfig:right_PIBA}
\includegraphics[width=0.25\linewidth]{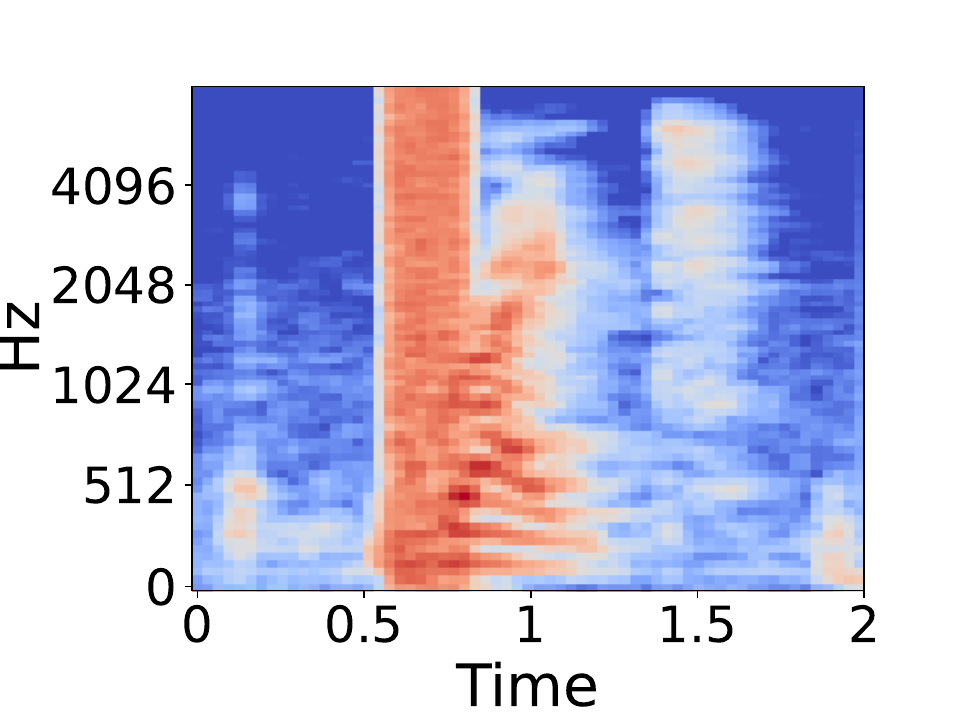}}
\subfloat[right\_DABA]{\label{fig:subfig:right_DABA}
\includegraphics[width=0.25\linewidth]{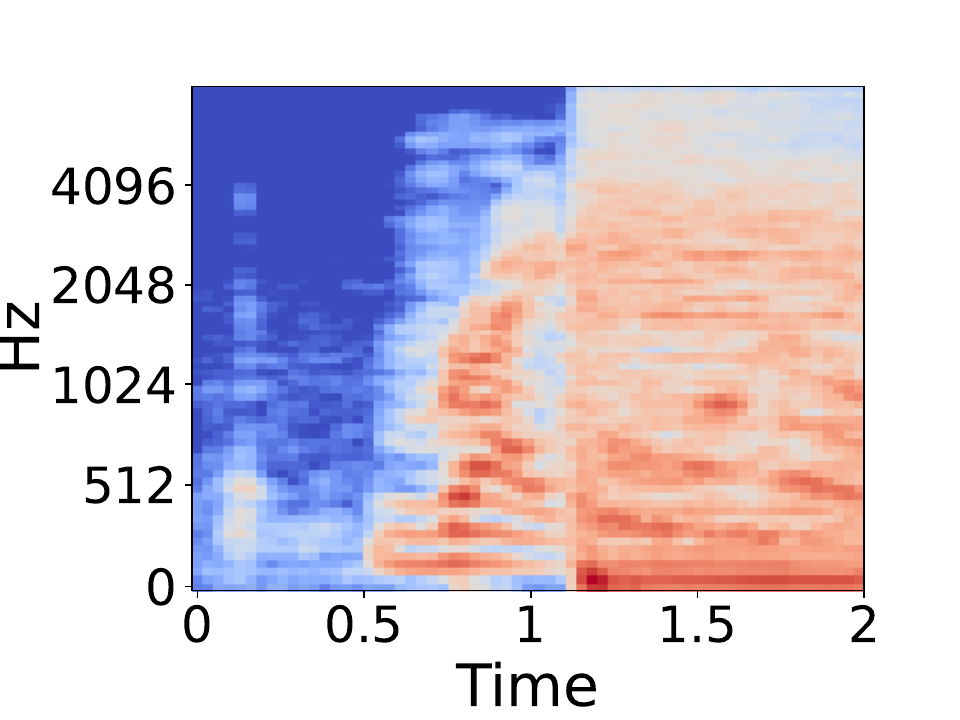}}
\subfloat[right\_Utralsonic]{\label{fig:subfig:right_Utral}
\includegraphics[width=0.25\linewidth]{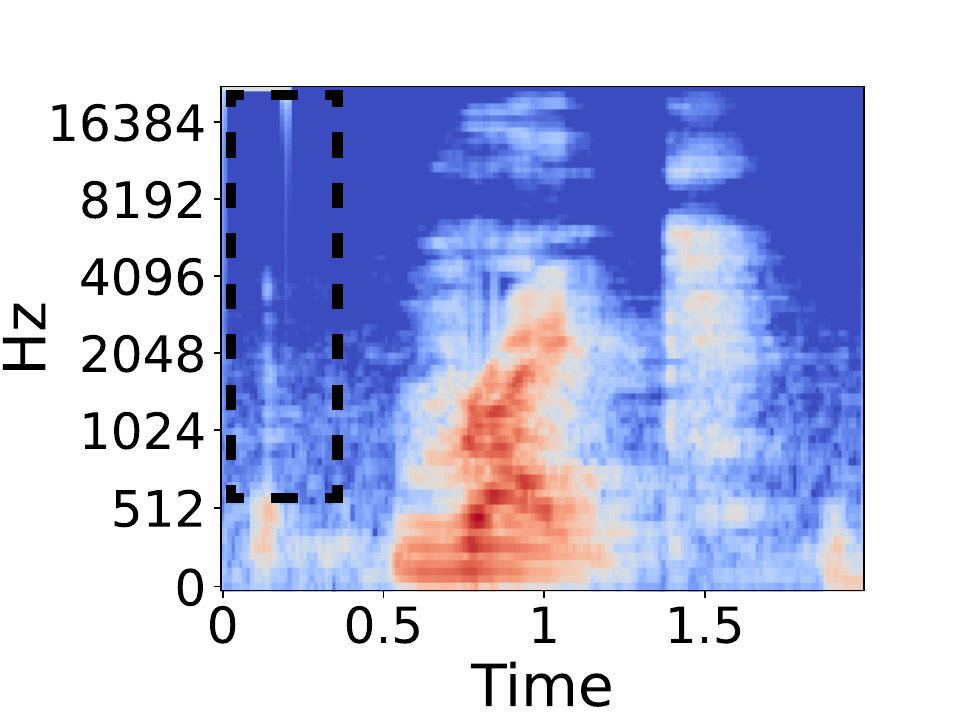}}
\vspace{-3.9mm}
\hfill
\subfloat[right\_JingleBack]{\label{fig:subfig:right_Style}
\includegraphics[width=0.25\linewidth]{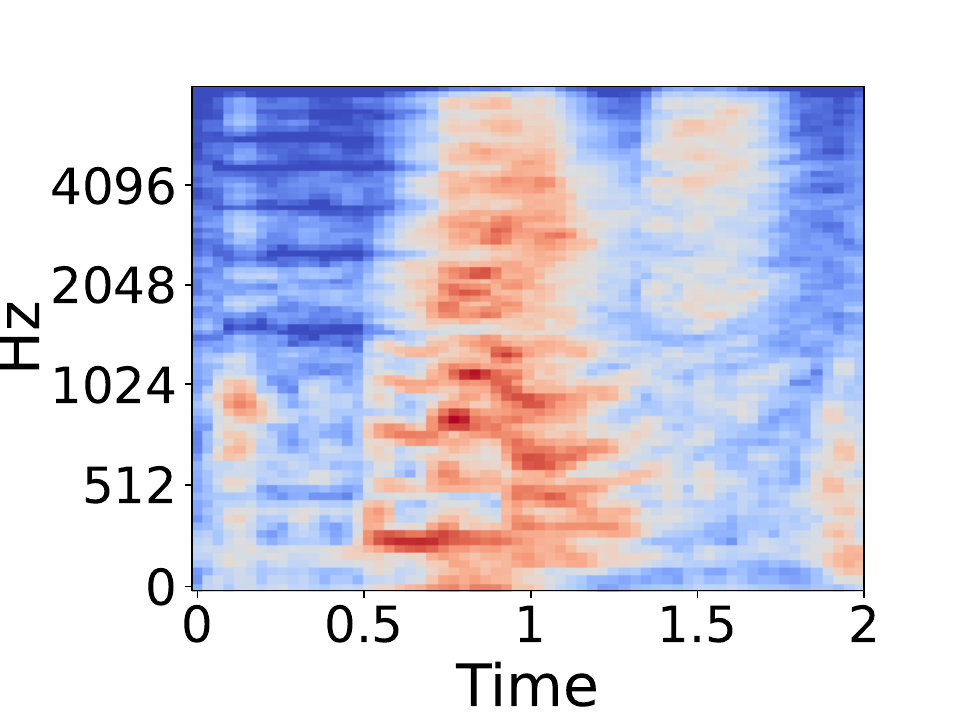}}
\subfloat[right\_Pitch]{\label{fig:subfig:right_Pitch}
\includegraphics[width=0.25\linewidth]{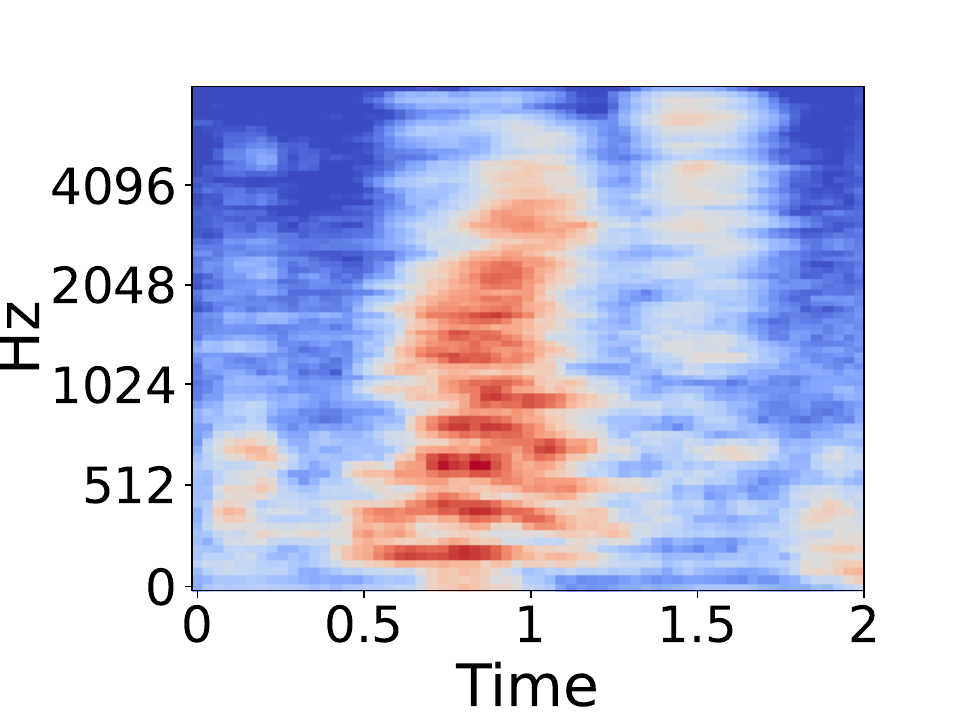}}
\subfloat[right\_PBSM]{\label{fig:subfig:right_PBSM}
\includegraphics[width=0.25\linewidth]{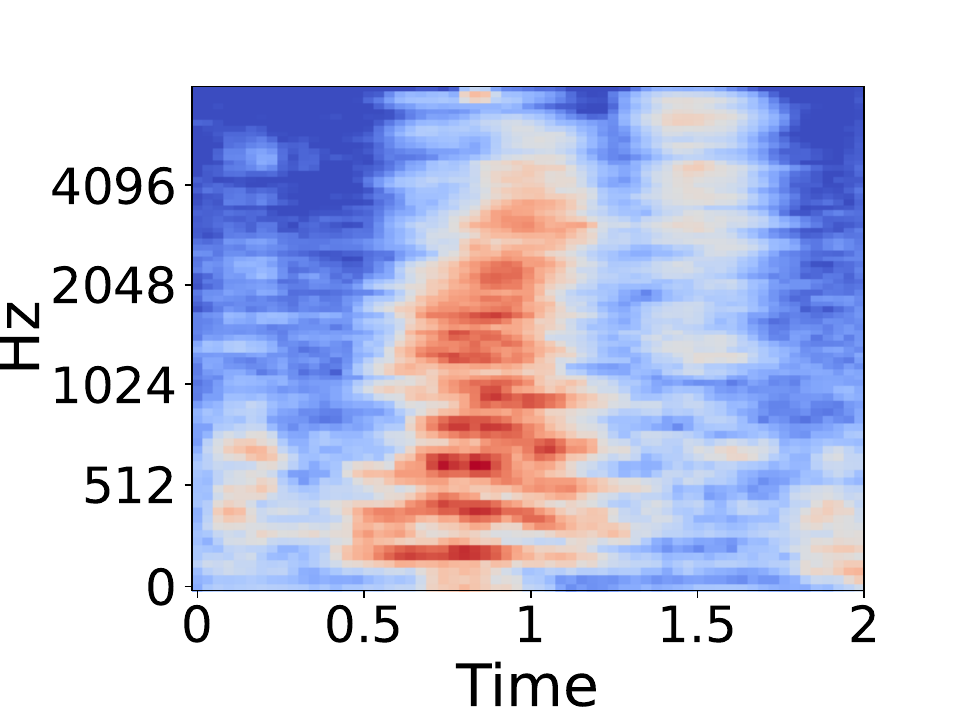}}
\subfloat[right\_VSVC]{\label{fig:subfig:right_VSVC}
\includegraphics[width=0.25\linewidth]{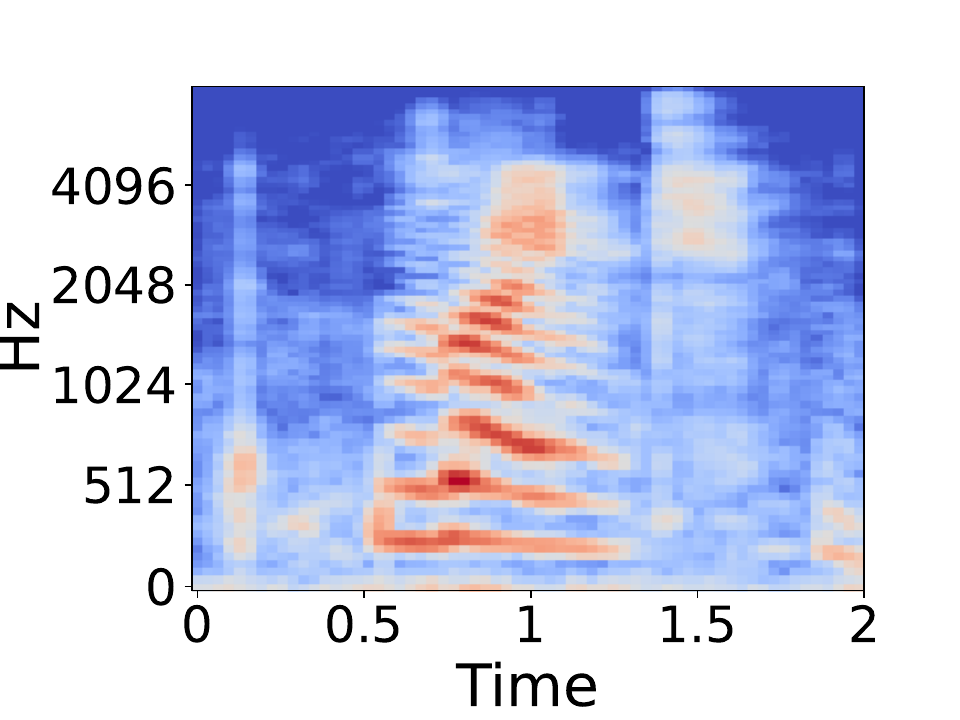}}

\caption{The spectrograms of different samples. In this example, we present the visualization of two benign audios (with the label `left' and `right') and their poisoned versions generated by different attacks. In particular, the black dashed box indicates the area where the person can easily identify the abnormality.}\label{fig:baseline_spec}

\end{figure*}

\vspace{0.3em}
\noindent \textbf{Attack Stealthiness.} We notice that the ASRs of baseline attacks (especially DABA and Ultrasonic) are higher than those of ours in some cases. However, it comes at the expense of stealthiness. As shown in Table~\ref{tab:human-validation}, the natural rates of all baseline attacks other than Ultrasonic are significantly lower than our PBSM and VSVC. For example, the natural rates of PIBA, DABA, and JingleBack are all 0\% while those of our PBSM and VSVC are near 100\%. Ultrasonic has a similar natural rate to that of benign samples simply because humans cannot hear ultrasound. However, it does not mean that this attack is stealthy. The victim users can still easily identify this attack by checking the spectrogram of samples (as shown in the area of the black dashed box in Figure~\ref{fig:subfig:left_Utral} and Figure~\ref{fig:subfig:right_Utral}). Users can also filter out ultrasonic trigger signals to depress this attack. These results verify the stealthiness of our attacks.

\vspace{0.3em}
In conclusion, our attacks can preserve high effectiveness while ensuring stealthiness. In contrast, existing baseline methods can be easily detected and defended.

\subsection{Ablation Study} 
\vspace{0.3em}
In this section, we discuss the effects of key parameters, including target label, poisoning rate, high-pitch signal, and timbre, of our PBSM and VSVC. We adopt SPC-10 as an example for our discussions. Unless otherwise specified, all settings are consistent to those stated in Section \ref{experimentalsetting}.

\begin{figure*}[!t]
\centering
\subfloat[LSTM]{\label{fig:subfig:poison_ablation_LSTM}
\includegraphics[width=0.24\linewidth]{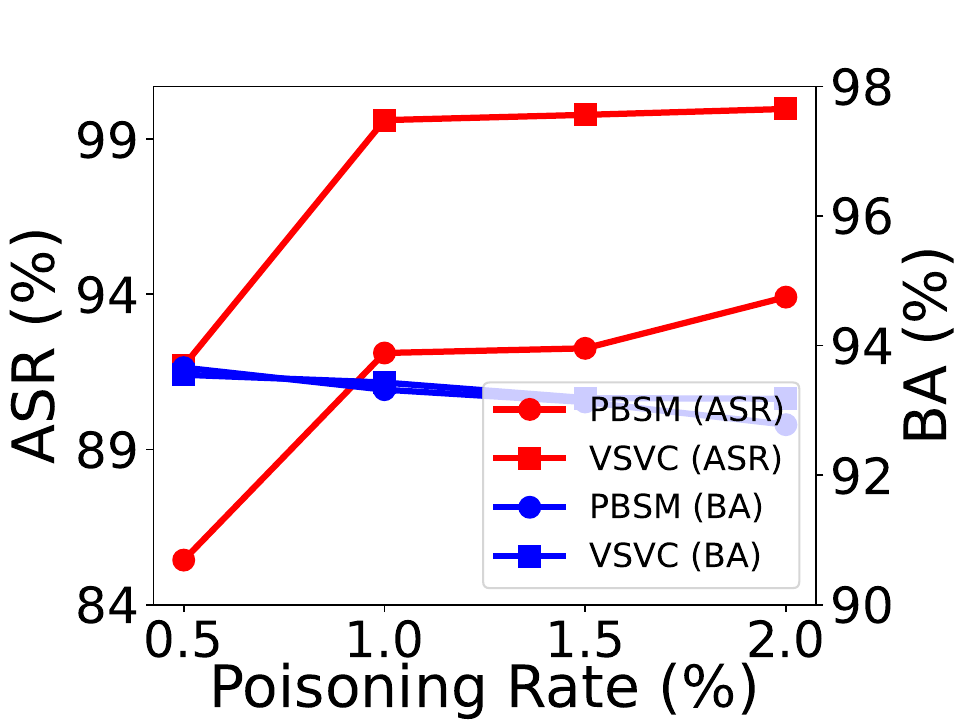}}
\subfloat[ResNet-18]{\label{fig:subfig:poison_ablation_ResNet-18}
\includegraphics[width=0.24\linewidth]{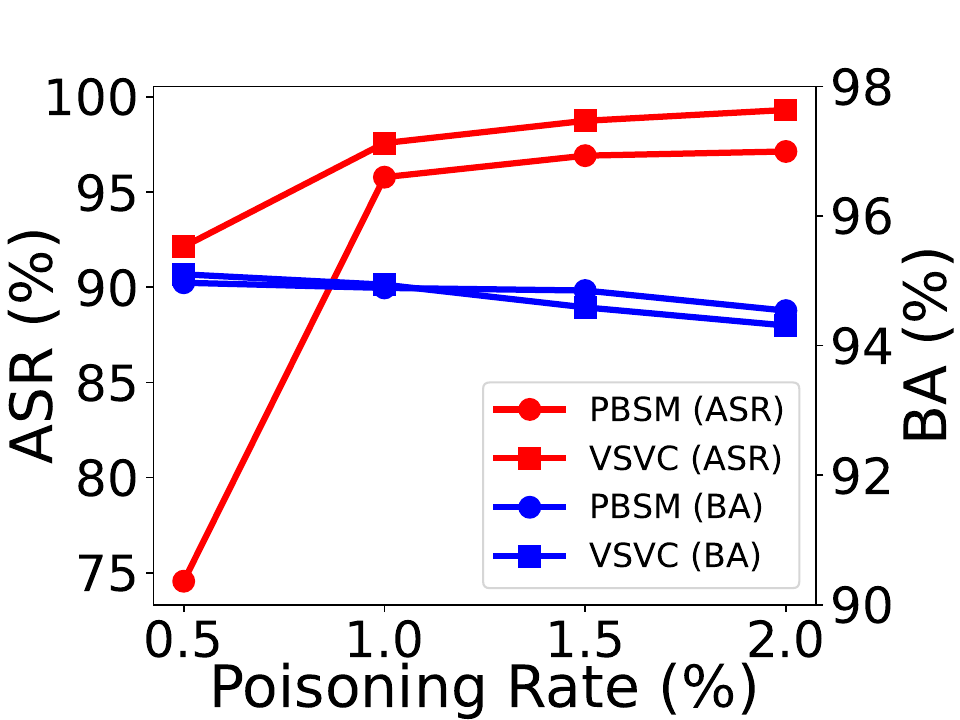}}
\subfloat[KWT]{\label{fig:subfig:poison_ablation_KWT}
\includegraphics[width=0.24\linewidth]{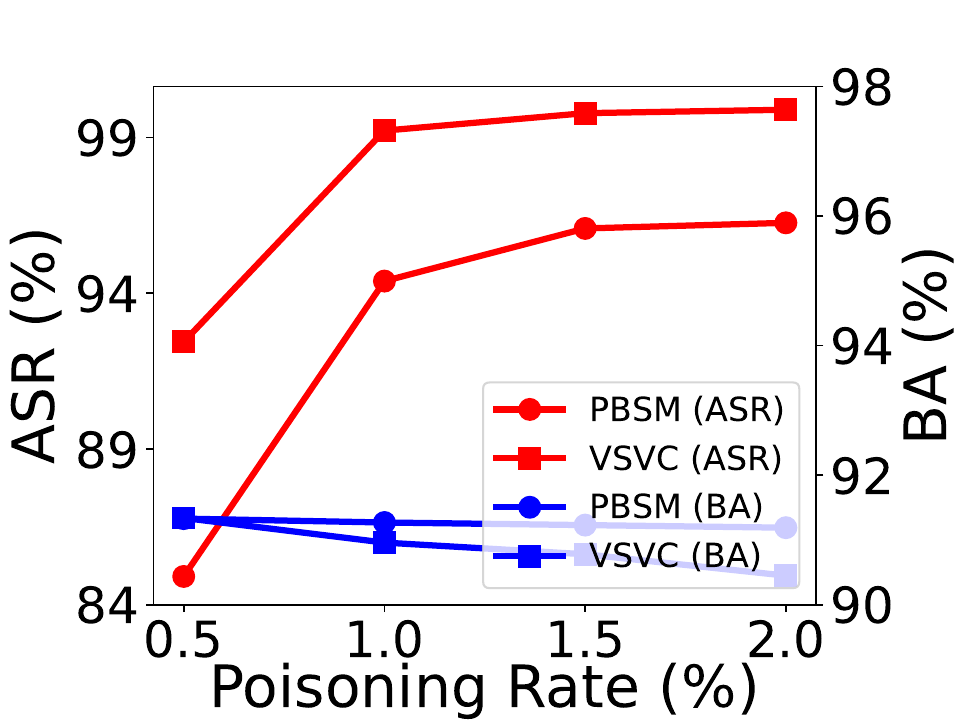}}
\subfloat[EAT]{\label{fig:subfig:poison_ablation_EAT}
\includegraphics[width=0.24\linewidth]{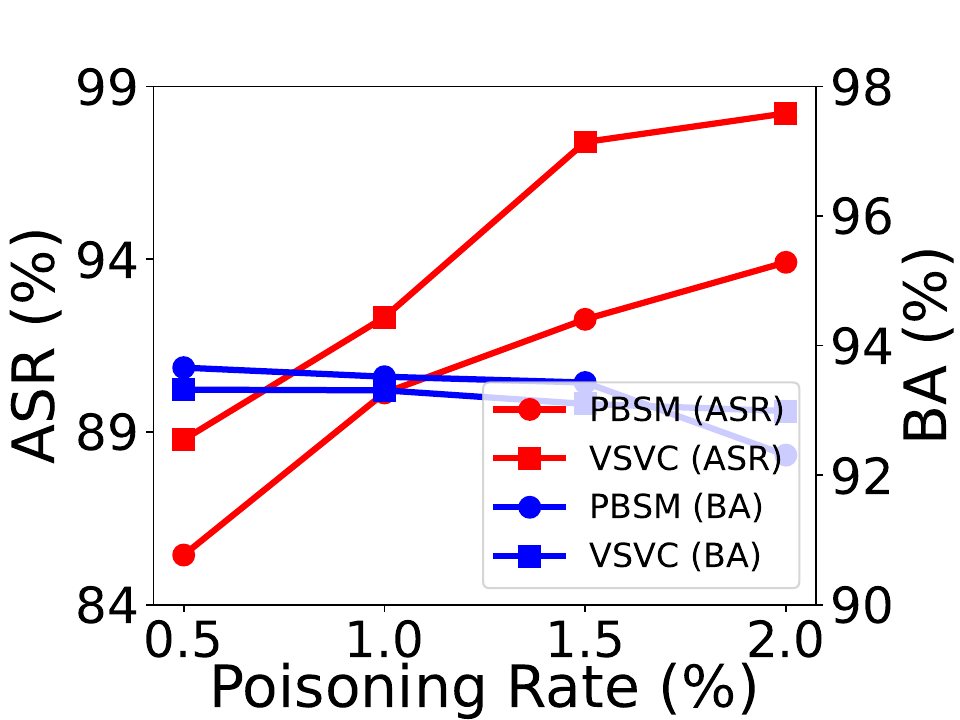}}
\caption{The performance of our PBSM and VSVC on the SPC-10 dataset under different poisoning rates.}\label{fig:poison_ablation}
%\vspace{-5mm}
\end{figure*}

\vspace{0.3em}
\noindent\textbf{Effects of the Poisoning Rate.} To explore the influences of the poisoning rate on our attacks, we conduct experiments with poisoning rates ranging from 0.5\% to 2.0\% against all four model structures. As shown in Figure~\ref{fig:poison_ablation}, the attack success rates (ASRs) of both PBSM and VSVC increase with the increase of the poisoning rate, although our attacks can reach promising attack performance by poisoning only 1\% training samples. However, the benign accuracy (BA) will decrease with the increase of the poisoning rates to some extent, \ie, there is a trade-off between ASR and BA. The adversaries should assign a suitable poisoning rate based on their needs.

\begin{table}[!ht]
\centering
\caption{The attack success rate (\%) $w.r.t.$ different boosted semitones on the SPC-10 dataset.}
\vspace{-0.5em}
\label{tab:pitch-ablation}
\resizebox{0.95\columnwidth}{!}{
\begin{tabular}{c|cccc}
\toprule
\tabincell{c}{Model$\rightarrow$\\ Semitone$\downarrow$} & LSTM  & ResNet-18 & KWT   & EAT   \\ \hline \rule{0pt}{8pt}
1             & 5.13  & 33.61     & 38.17 & 37.91 \\ \hline \rule{0pt}{8pt}
3             & 70.61 & 69.70     & 79.08 & 46.17 \\ \hline \rule{0pt}{8pt}
5             & 80.74 & 85.65     & 82.13 & 73.09 \\ \hline \rule{0pt}{8pt}
7             & 86.09 & 89.35     & 83.19 & 81.61 \\ \bottomrule
\end{tabular}}
\end{table}

\begin{figure}[!t]
\vspace{-1em}
\centering
\subfloat[PBSM]{\label{fig:subfig:PBSM_label_ablation}
\includegraphics[width=0.473\columnwidth]{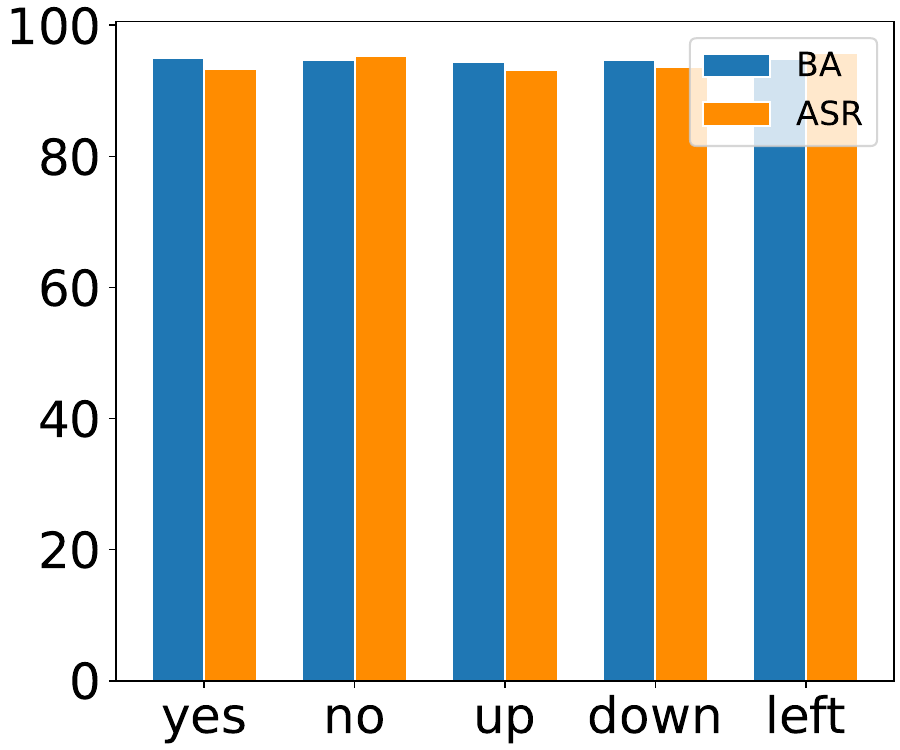}}
\subfloat[VSVC]{\label{fig:subfig:VSVC_label_ablation}
\includegraphics[width=0.473\columnwidth]{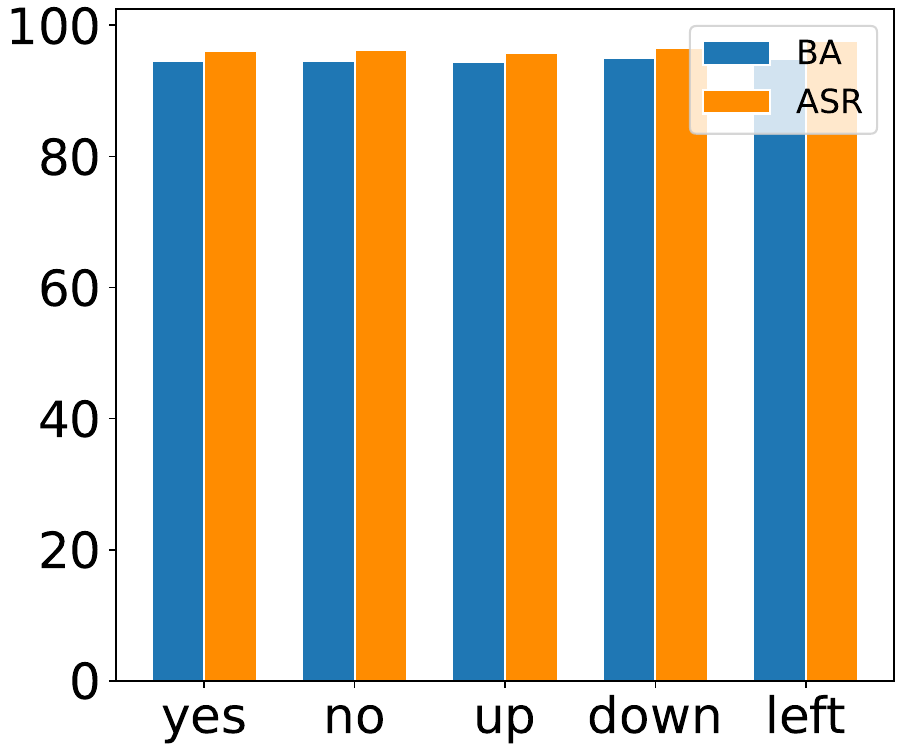}}
\caption{The effects of the target label on our PBSM and VSVC attacks on the SPC-10 dataset.}\label{fig:label_ablation}
\vspace{-0.5em}
\end{figure}

\vspace{0.3em}
\noindent\textbf{Effects of the Target Label.} To verify that our PBSM and VSVC are still effective under different target labels, we conduct experiments with ResNet. As shown in Figure~\ref{fig:label_ablation}, the attack success rates of both PBSM and VSVC are similar across all evaluated target labels. Specifically, the ASRs are larger than 93\% in all cases, while the decrease of benign accuracy compared to 'no attack' is less than 1\%. These results show that target labels have minor effects on our attacks. The adversaries can select any target class based on their needs.

\begin{figure*}[!t]
\centering
\subfloat[left (Benign)]{
\includegraphics[width=0.19\linewidth]{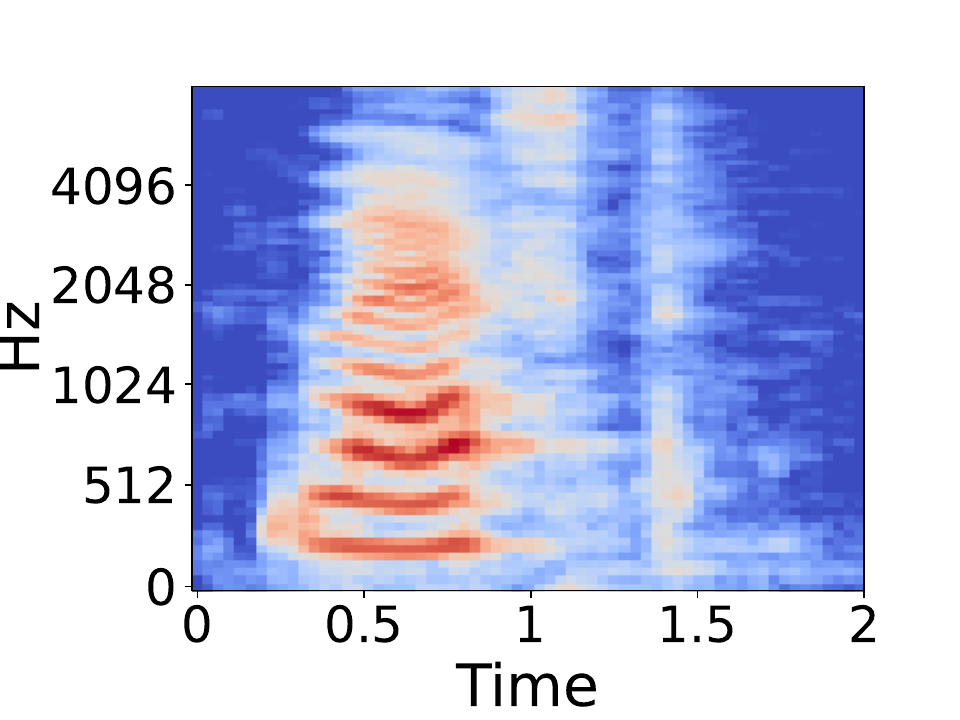}}
\subfloat[left (1 Semitone)]{
\includegraphics[width=0.19\linewidth]{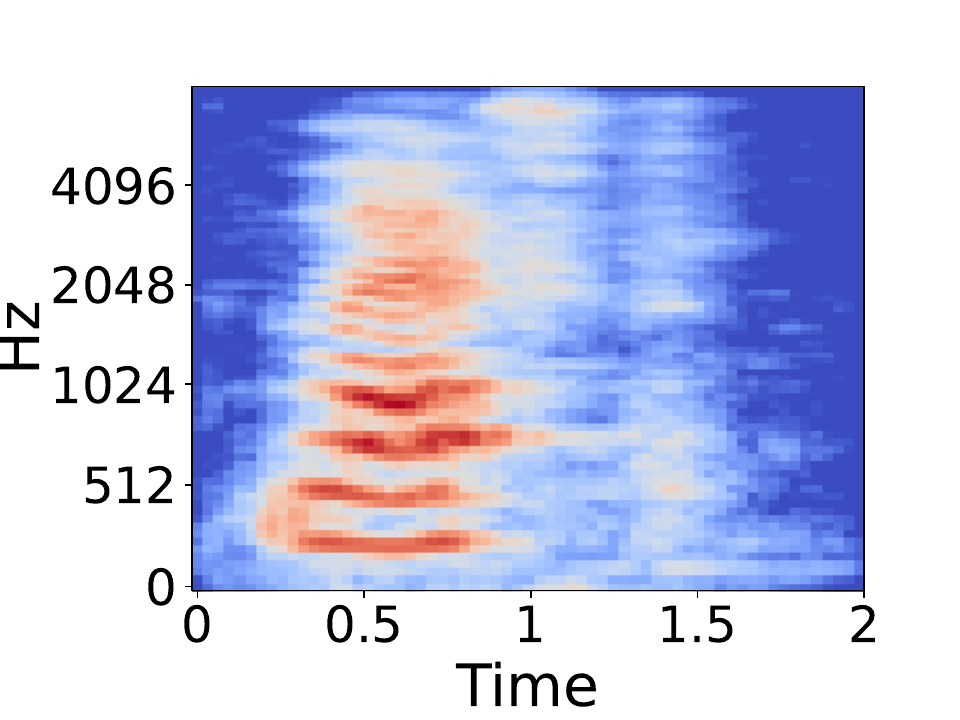}}
\subfloat[left (3 Semitones)]{
\includegraphics[width=0.19\linewidth]{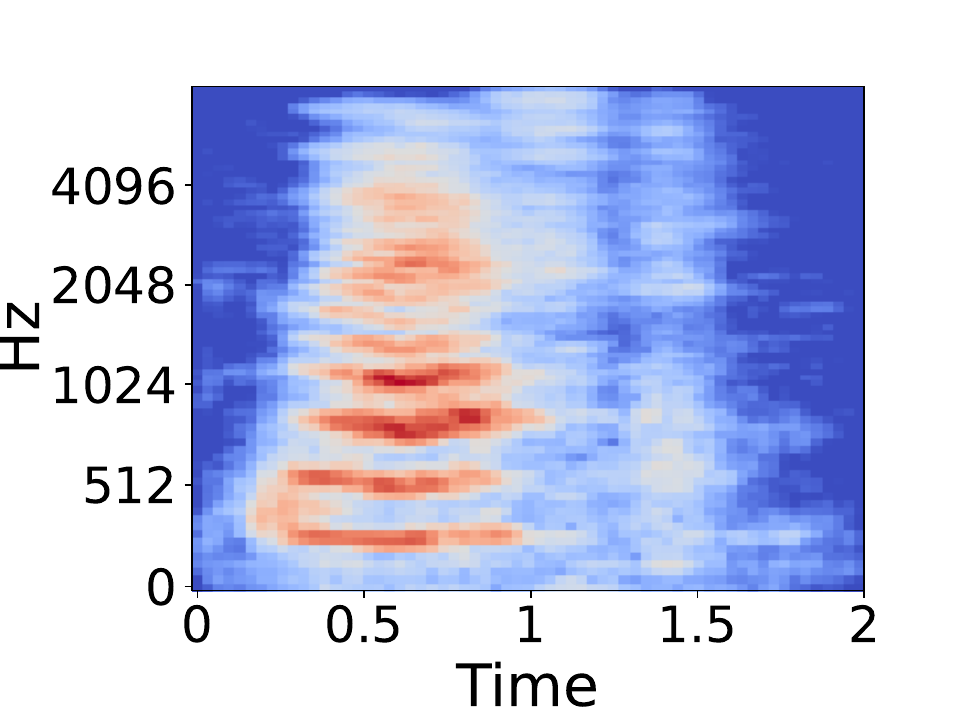}}
\subfloat[left (5 Semitones)]{
\includegraphics[width=0.19\linewidth]{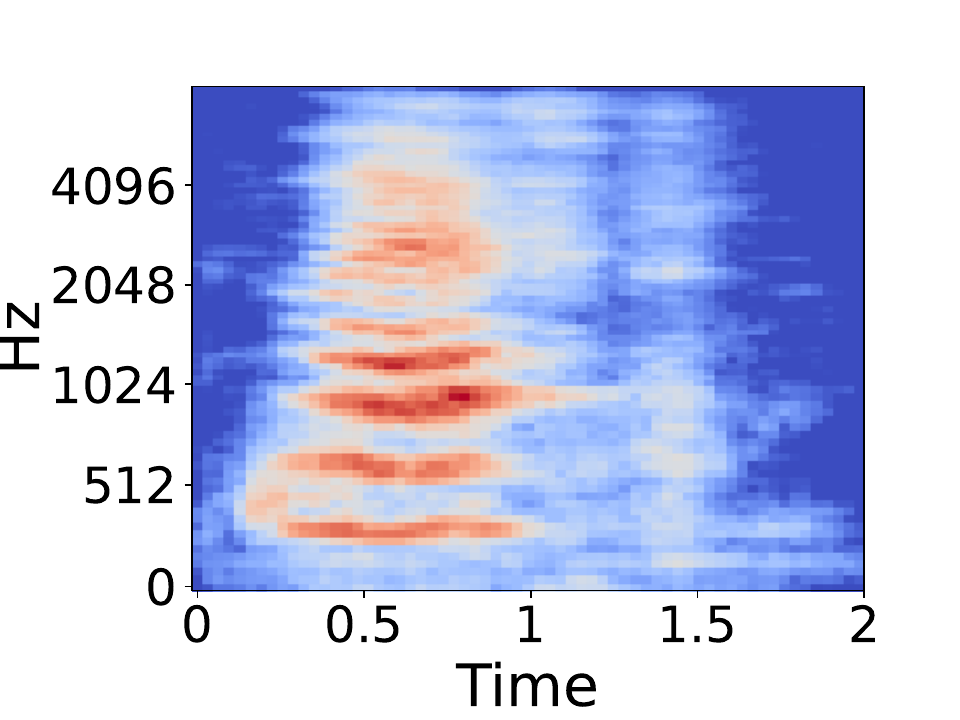}}
\subfloat[left (7 Semitones)]{
\includegraphics[width=0.19\linewidth]{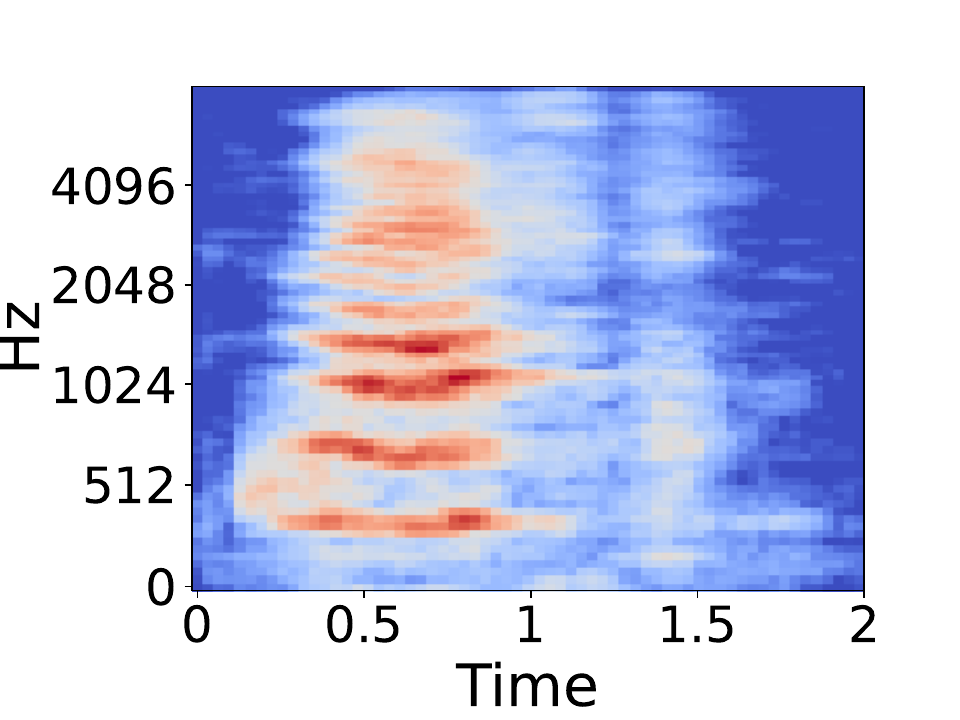}}
\hfill
\vspace{-4mm}
\subfloat[right (Benign)]{
\includegraphics[width=0.19\linewidth]{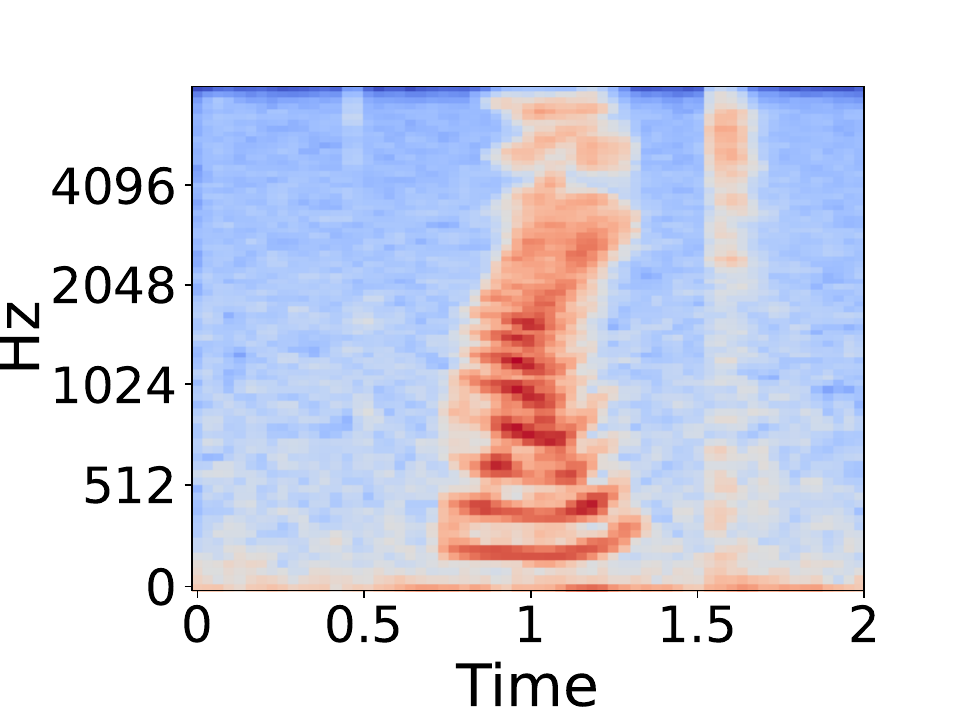}}
\subfloat[right (1 Semitone)]{
\includegraphics[width=0.19\linewidth]{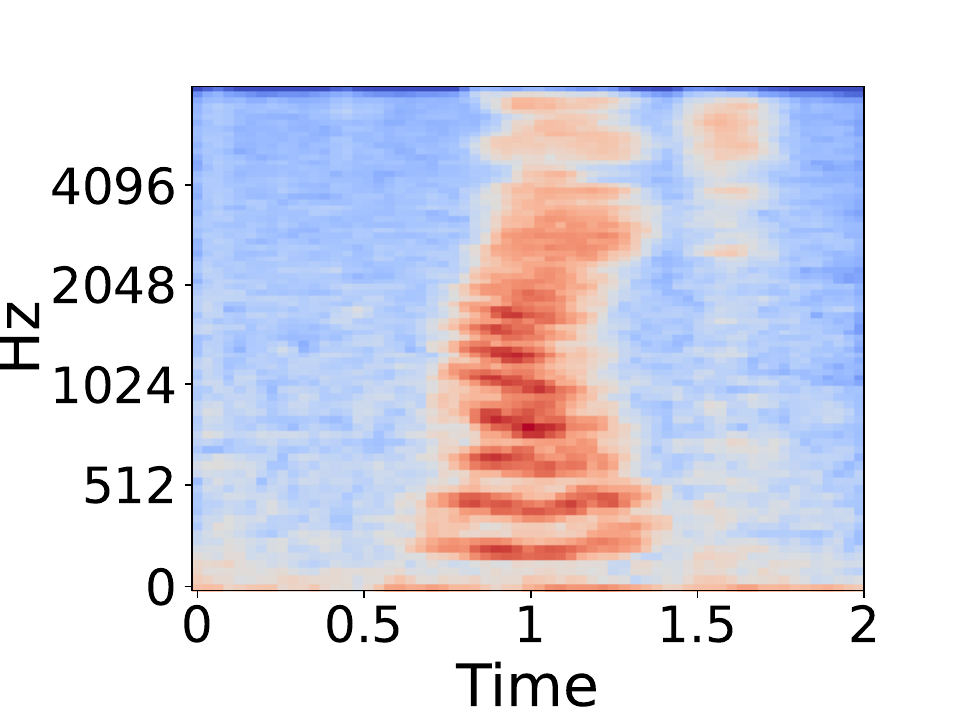}}
\subfloat[right (3 Semitones)]{
\includegraphics[width=0.19\linewidth]{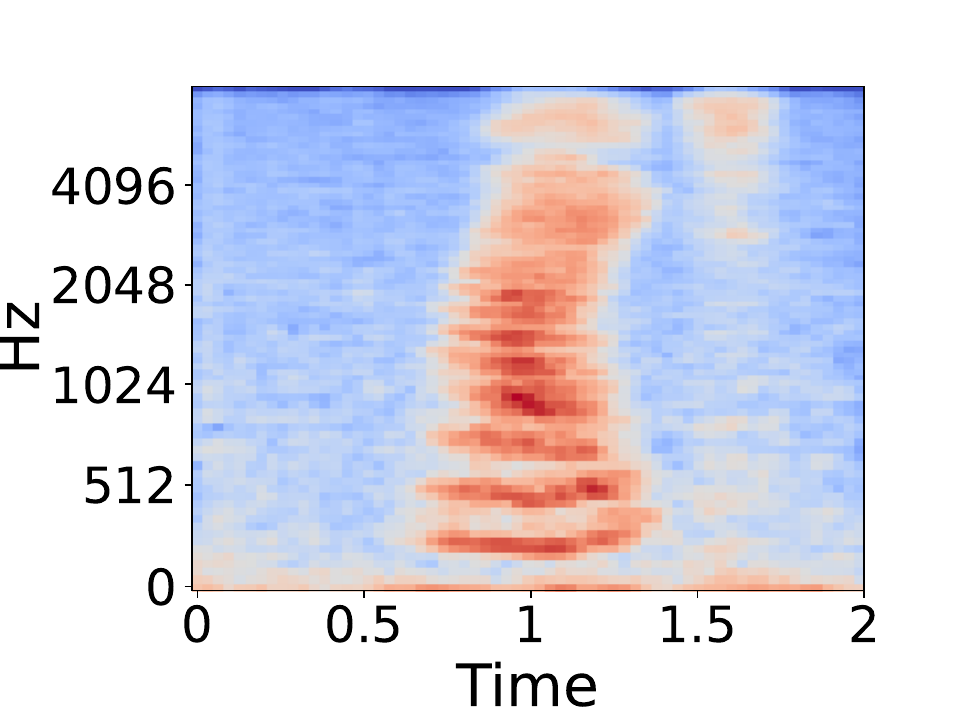}}
\subfloat[right (5 Semitones)]{
\includegraphics[width=0.19\linewidth]{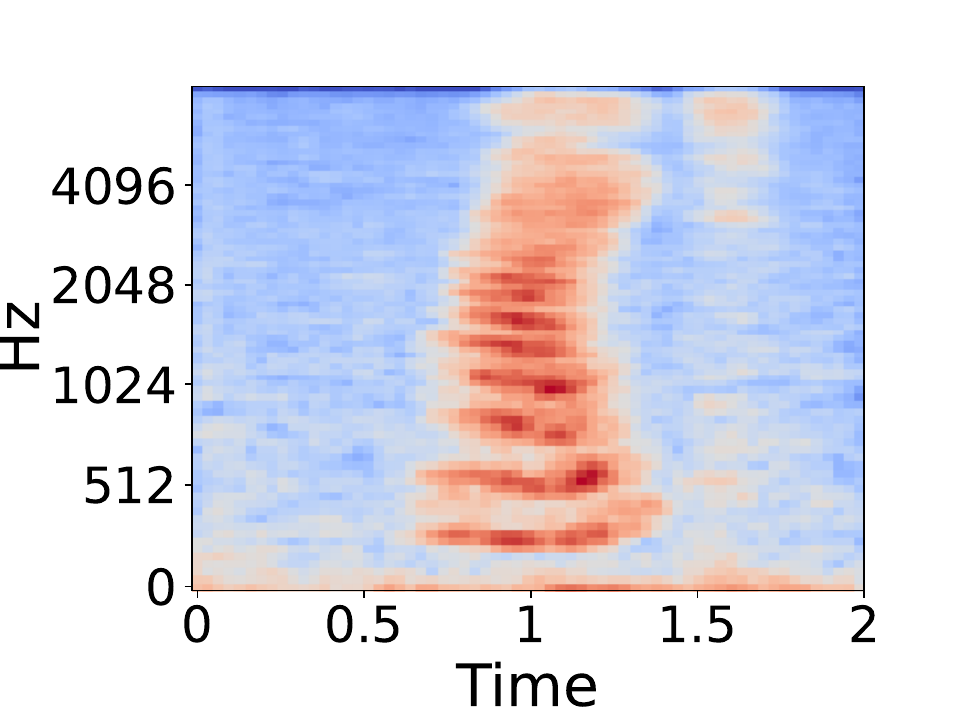}}
\subfloat[right (7 Semitones)]{
\includegraphics[width=0.19\linewidth]{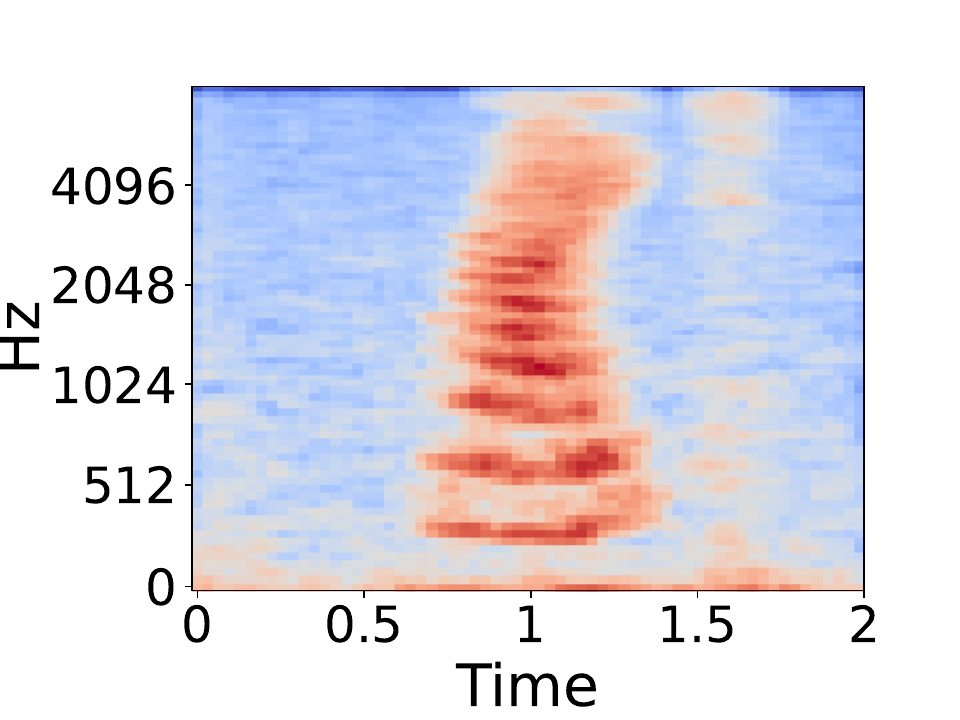}}
\caption{The spectrograms of samples whose pitch is boosted with different semitone. In this example, we present the visualization of two benign audios (with the label `left' and `right') and their boosted versions.}
%\vspace{-5mm}
\label{fig:pitch-boosted}
\end{figure*}

\vspace{0.3em}
\noindent\textbf{Effects of the Pitch Boosting.} In this part, we show that pitch boosting used in our PBSM itself can serve as the pitch-type trigger and explore its effects. Specifically, we increase the pitch range from one semitone to seven semitones and evaluate the attack success rate (ASR). The example of the spectrograms of samples with different boosted semitone is show in Figure \ref{fig:pitch-boosted}. As shown in Table~\ref{tab:pitch-ablation}, the ASR increases with the increase of semitones, as we expected. Specifically, the ASRs are larger than 80\% in three out of all four cases when we boost five semitones. However, we have to notice that excessive pitch boosting can lead to significant sound distortion and therefore decreasing attack stealthiness.

\vspace{0.3em}
\noindent\textbf{Effects of the Short-duration High-pitch Signal.} To verify that inserting a high-pitch signal is critical for our PBSM, we compare its attack success rate to that of its pitch-only variant where we only increase the pitch without adding the high-pitch signal. As shown in Table~\ref{tab:PBSM_ablation}, although the pitch-only method can have some attack effects, introducing high-pitch signal can significantly improve the attack effectiveness. Specifically, the attack success rates of PBSM is 10\% higher than that of its pitch-only variant in all cases. These results verify the effectiveness of our PBSM.

\begin{table}[!t]
\centering
\caption{The attack success rate (\%) of pitch-only attack and PBSM attack on the SPC-10 dataset.}
\vspace{-0.5em}
\label{tab:PBSM_ablation}
\resizebox{\columnwidth}{!}{
\begin{tabular}{c|cccc}
\toprule
Method$\downarrow$, Model$\rightarrow$ & LSTM  & ResNet-18 & KWT   & EAT   \\ \hline \rule{0pt}{8pt}
Pitch-Only                             & 80.74 & 85.65    & 82.13 & 73.09 \\ \hline \rule{0pt}{8pt}
PBSM                                   & \textbf{92.11} & \textbf{95.78}    & \textbf{94.39} & \textbf{90.13} \\ \bottomrule
\end{tabular}}
\end{table}

\begin{figure*}[htbp]
\centering
\subfloat[left (Benign)]{
\includegraphics[width=0.2\linewidth]{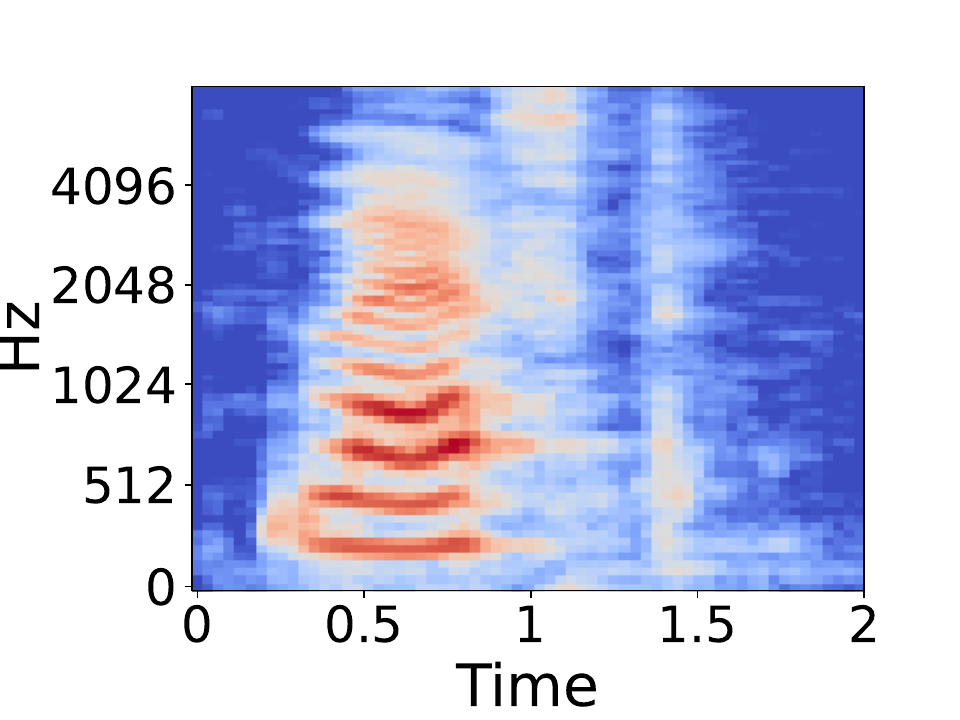}}
\subfloat[left (timbre\_a)]{
\includegraphics[width=0.2\linewidth]{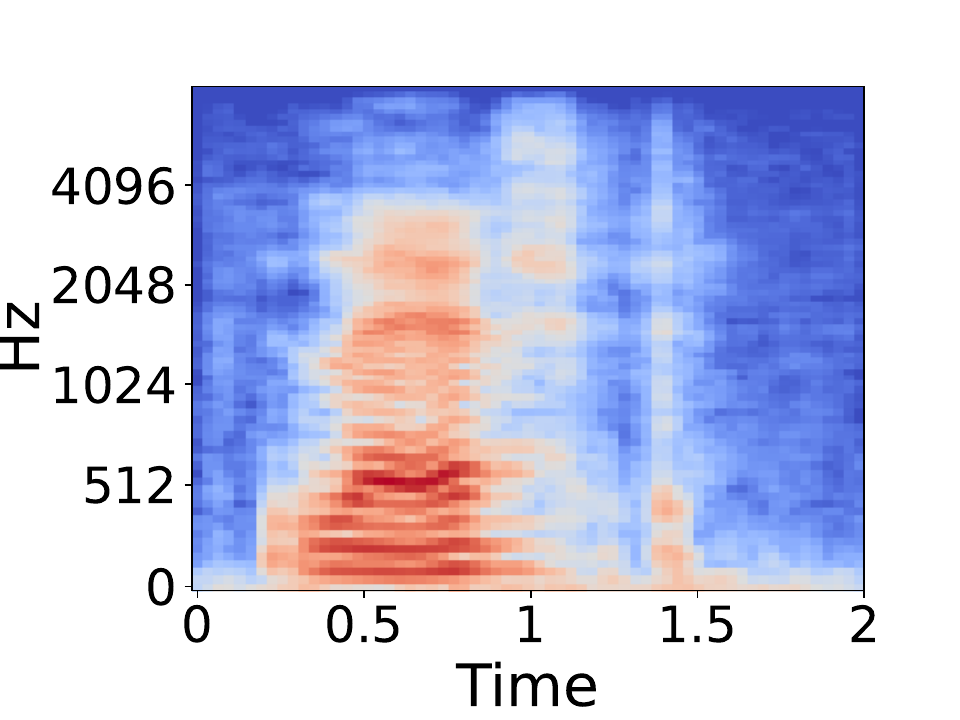}}
\subfloat[left (timbre\_b)]{
\includegraphics[width=0.2\linewidth]{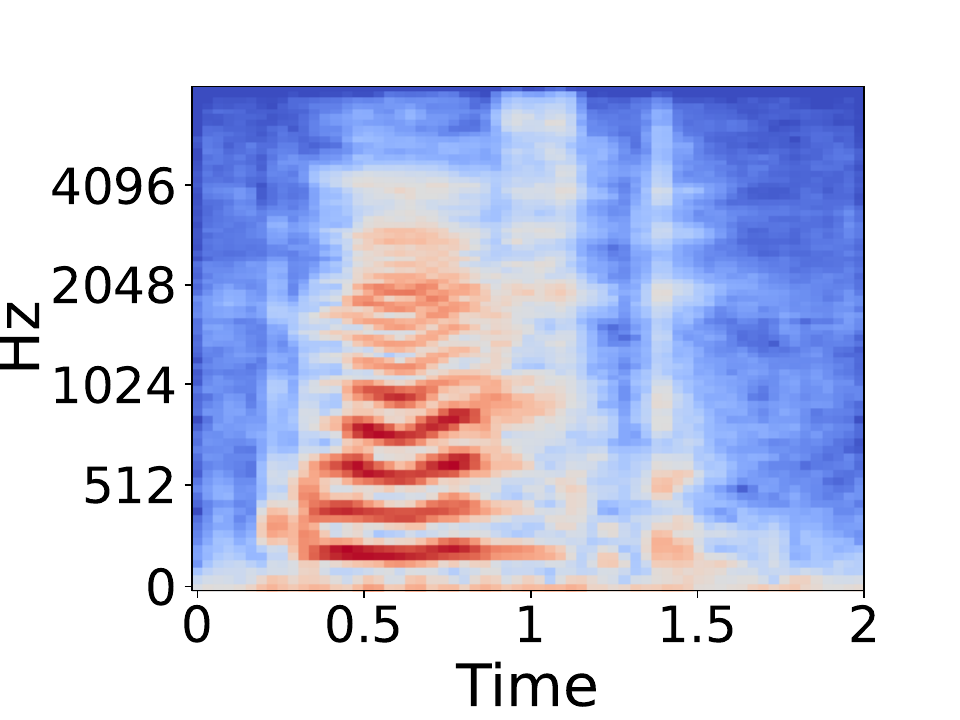}}
\subfloat[left (timbre\_c)]{
\includegraphics[width=0.2\linewidth]{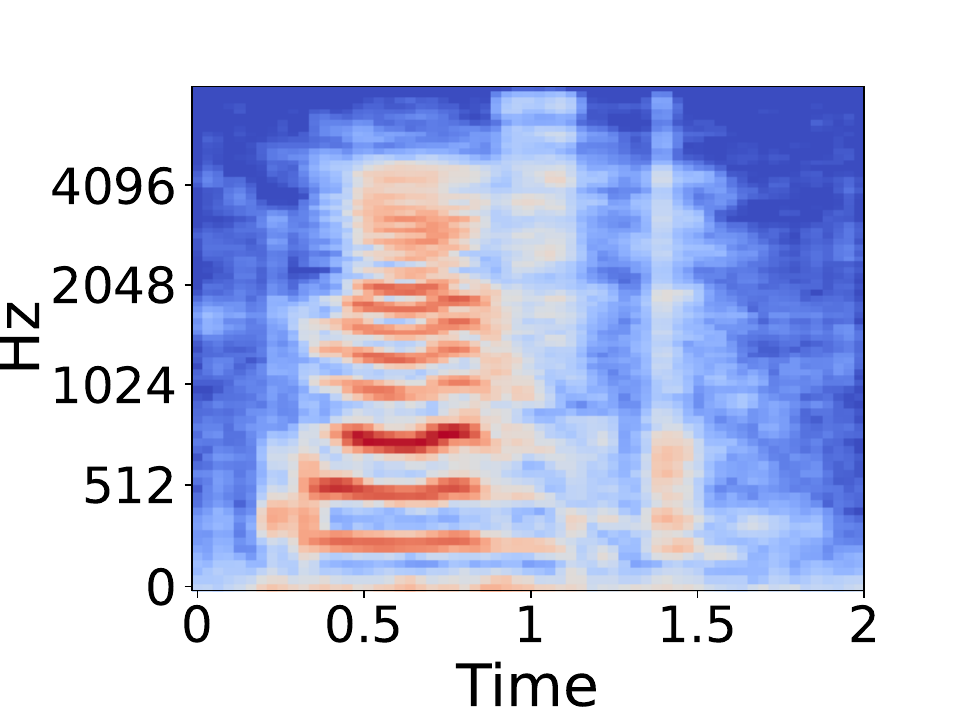}}
\subfloat[left (timbre\_d)]{
\includegraphics[width=0.2\linewidth]{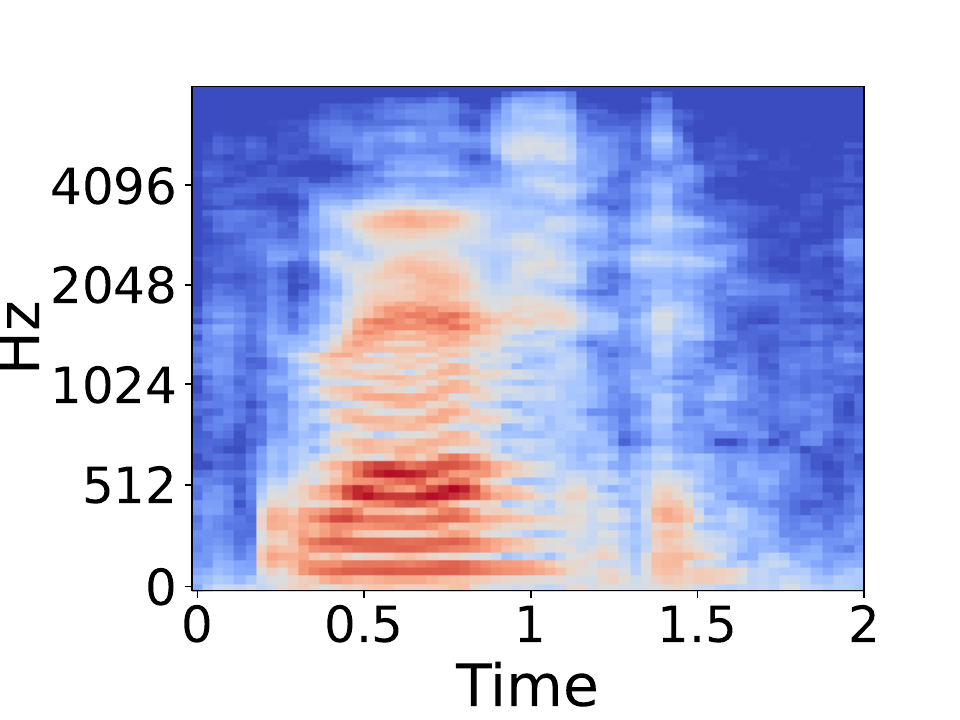}}
\hfill
\vspace{-4mm}
\subfloat[right (Benign)]{
\includegraphics[width=0.2\linewidth]{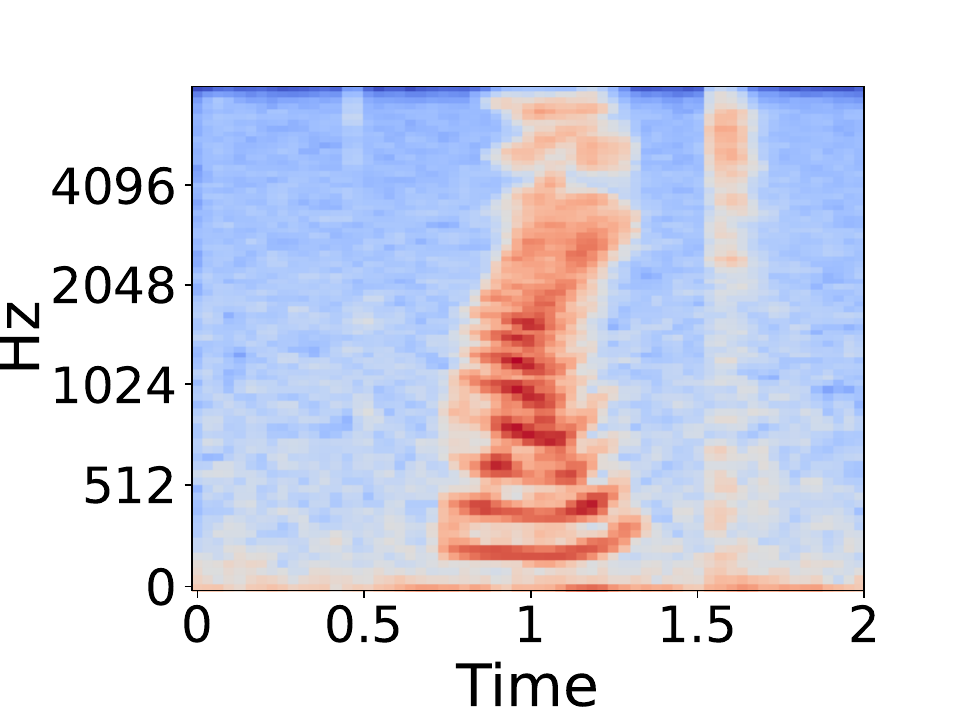}}
\subfloat[right (timbre\_a)]{
\includegraphics[width=0.2\linewidth]{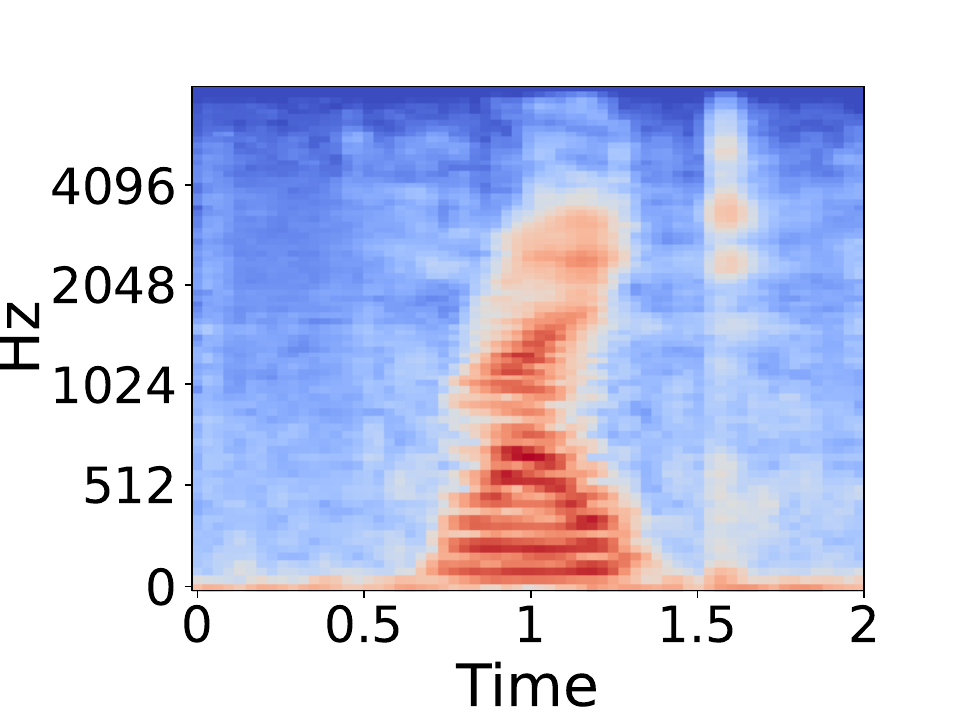}}
\subfloat[right (timbre\_b)]{
\includegraphics[width=0.2\linewidth]{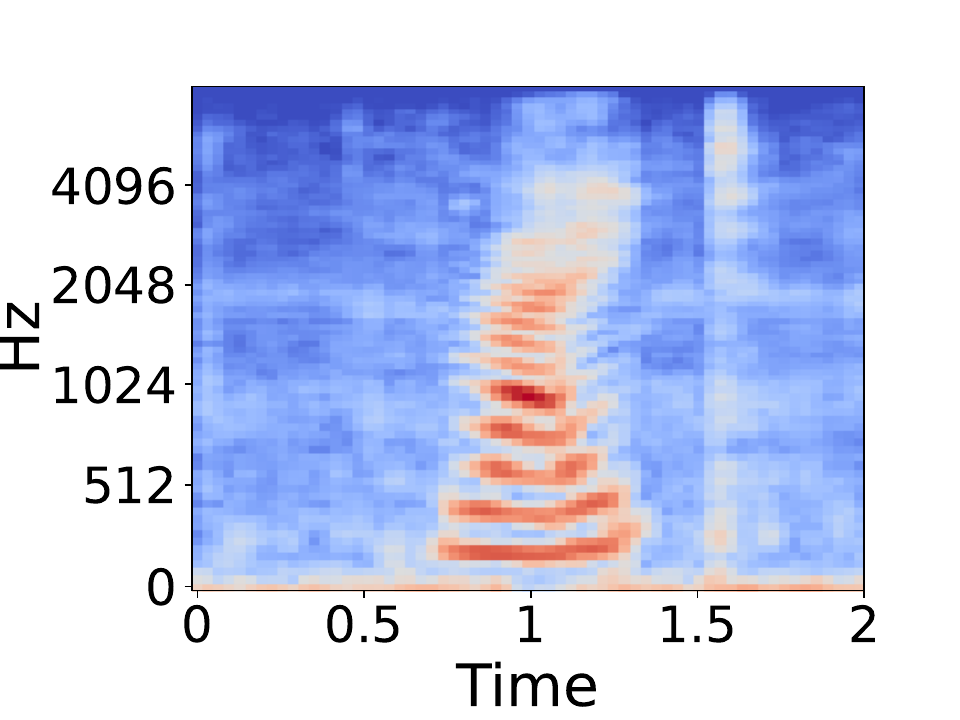}}
\subfloat[right (timbre\_c)]{
\includegraphics[width=0.2\linewidth]{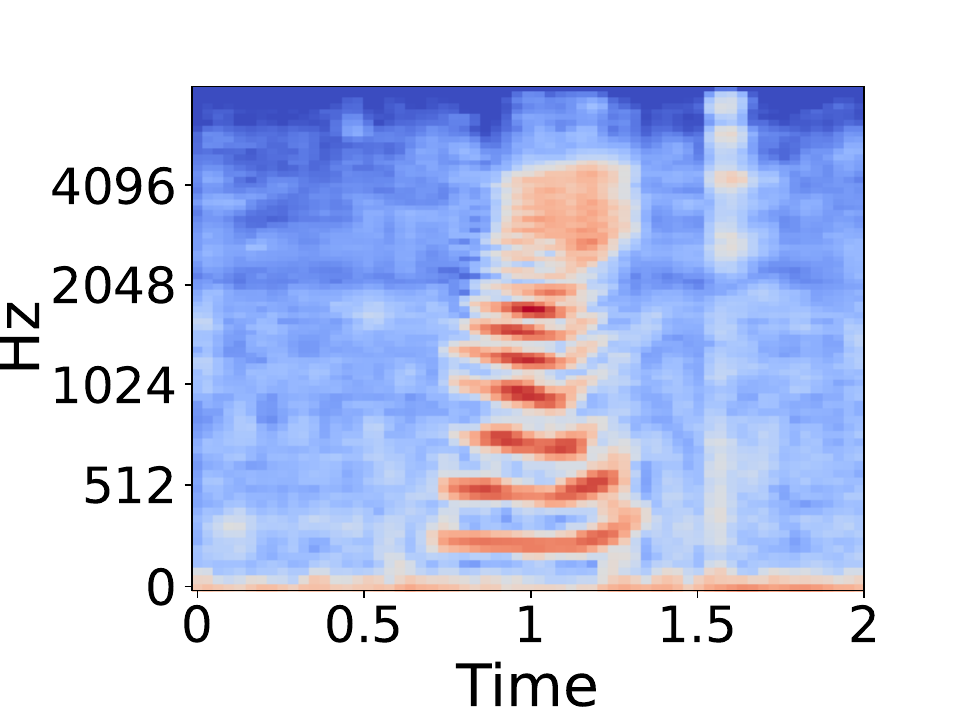}}
\subfloat[right (timbre\_d)]{
\includegraphics[width=0.2\linewidth]{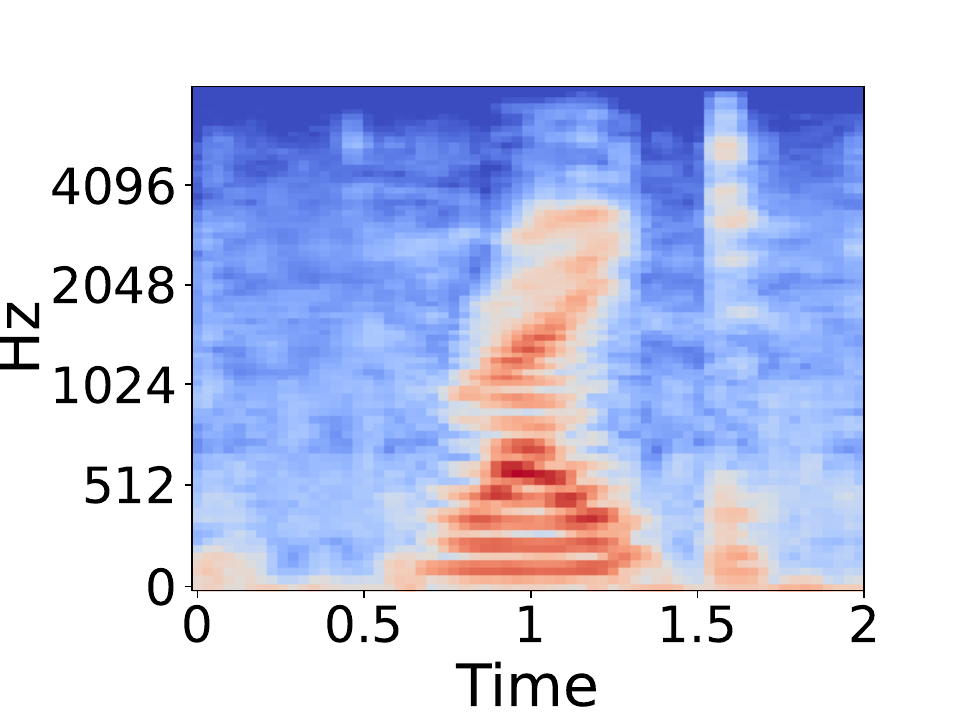}}
\caption{The spectrograms of samples with different timbre. In this example, we present the visualization of two benign audios (with the label 'left' and 'right') and their variants with different timbres.}
\label{different-timbre-specrograms}
%\vspace{-5mm}
\end{figure*}

\vspace{0.3em}
\noindent\textbf{Effects of the Timbre.} To verify that our VSVC is still effective with different timbres, we conducted experiments on the SPC-10 dataset. The example of the spectrograms of samples with different timbres is show in Figure \ref{different-timbre-specrograms}. As shown in Table~\ref{tab:different_timbres}, the ASRs of VSVC are similar across all evaluated timbres. Specifically, the ASRs are larger than 91\% in all cases, while the decrease of benign accuracy compared to 'no attack' is only about 1\%. These results indicate that timbre selection has mild effects on our attack. The adversaries can select any timbre based on their needs.

% \usepackage{multirow}
\begin{table}[!t]
\centering
\caption{The attack success rate (\%) of our VSVC attack with different timbres on the SPC-10 dataset.}
\vspace{-0.5em}
\label{tab:different_timbres}
\resizebox{\columnwidth}{!}{
\begin{tabular}{c|c|cccc}
\toprule
\multicolumn{1}{c|}{Timbre$\downarrow$}                     & \tabincell{c}{Metric$\downarrow$ \\ Model$\rightarrow$}                                                             & \multicolumn{1}{c}{LSTM}   & \multicolumn{1}{c}{ResNet-18}   & \multicolumn{1}{c}{KWT}  & \multicolumn{1}{c}{EAT} \\ \hline 
\multicolumn{1}{c|}{ \rule{0pt}{8pt} \multirow{2}{*}{(a)}} & \multirow{2}{*}{\begin{tabular}[c]{@{}c@{}}BA (\%)\\ ASR (\%)\end{tabular}} & \multicolumn{1}{c}{93.56} & \multicolumn{1}{c}{94.88} & \multicolumn{1}{c}{91.04} & \multicolumn{1}{c}{93.13}    \\
\multicolumn{1}{c|}{}                           &                                                                           & \multicolumn{1}{c}{98.52} & \multicolumn{1}{c}{97.51} & \multicolumn{1}{c}{98.71} & \multicolumn{1}{c}{91.33}    \\ \hline
\multicolumn{1}{c|}{ \rule{0pt}{8pt} \multirow{2}{*}{(b)}} & \multirow{2}{*}{\begin{tabular}[c]{@{}c@{}}BA (\%)\\ ASR (\%)\end{tabular}} & \multicolumn{1}{c}{93.32} & \multicolumn{1}{c}{94.76} & \multicolumn{1}{c}{91.36} & \multicolumn{1}{c}{93.21}    \\
\multicolumn{1}{c|}{}                           &                                                                           & 99.08                     & 98.53                     & 98.81                     & 93.11                        \\ \hline \rule{0pt}{8pt}
\multirow{2}{*}{(c)}                      & \multirow{2}{*}{\begin{tabular}[c]{@{}c@{}}BA (\%)\\ ASR (\%)\end{tabular}} & 92.88                     & 94.23                     & 90.98                     & 92.89                        \\
                                                &                                                                           & 97.60                     & 96.65                     & 97.87                     & 92.30                        \\ \hline \rule{0pt}{8pt}
\multirow{2}{*}{(d)}                      & \multirow{2}{*}{\begin{tabular}[c]{@{}c@{}}BA (\%)\\ ASR (\%)\end{tabular}} & 93.15                     & 94.22                     & 90.77                     & 92.78                        \\
                                                &                                                                           & 98.15                     & 96.73                     & 98.69                     & 92.14                        \\ \hline \rule{0pt}{8pt}
\multirow{2}{*}{(e)}                      & \multirow{2}{*}{\begin{tabular}[c]{@{}c@{}}BA (\%)\\ ASR (\%)\end{tabular}} & 92.61                     & 94.35                     & 91.33                     & 92.39                        \\
                                                &                                                                           & 99.17                     & 98.92                     & 99.08                     & 94.47                        \\ \bottomrule
\end{tabular}}
\end{table}

\begin{table}[!t]
\centering
\caption{The performance of VSVC without and with voiceprint selection under the multiple-backdoor setting.}
\vspace{-0.5em}
\label{tab:Multi_target_result}
\resizebox{\columnwidth}{!}{
\begin{tabular}{c|c|cccc}
\toprule
Method$\downarrow$                         & Metric$\downarrow$, Model$\rightarrow$                                                             & LSTM   & ResNet-18   & KWT   & EAT \\ \hline \rule{0pt}{8pt}
\multirow{2}{*}{VSVC (w/o)} & \multirow{2}{*}{\begin{tabular}[c]{@{}c@{}}BA (\%)\\ ASR (\%)\end{tabular}} & 91.23 & 94.58 & 88.54  & 91.43    \\
                               &                                                                           & 89.10 & 91.24 & 92.34  & 87.65    \\ \hline \rule{0pt}{8pt}
\multirow{2}{*}{VSVC (w/)}          & \multirow{2}{*}{\begin{tabular}[c]{@{}c@{}}BA (\%)\\ ASR (\%)\end{tabular}} & 92.05 & 95.05 & 90.13  & 93.14    \\
                               &                                                                           & 92.77 & 97.78 & 97.03  & 93.78    \\ 
\bottomrule
\end{tabular}}
\end{table}

\vspace{0.3em}
\noindent\textbf{Effects of the Voiceprint Selection.}\label{sec:mutitarget-attack} To verify that voiceprint selection is critical for our VSVC under the multi-backdoor setting, we compare its attack success rate to that of its variant where we randomly select timbre candidates for voice conversion. In these experiments, we select three timbre candidates for discussions. As shown in Table~\ref{tab:Multi_target_result}, although the random selection variant can also have some attack effects, the introduction of voiceprint selection can significantly improve attack effectiveness. Specifically, the attack success rates of VSVC are 5\% higher than those of its random selection variant in almost all cases. These results verify the effectiveness of the voiceprint selection introduce in our VSVC.

\subsection{The Resistance to Potential Defenses}
Currently, there are many backdoor defenses designed to reduce backdoor threats in image classification tasks \cite{guo2023scale,xiang2023umd,jebreel2023defending}. However, most of them cannot be directly used in audio tasks since they are specified for the image domain. Accordingly, in this paper, we evaluate our attacks under three classical and representative cross-domain defenses, including model pruning \cite{fine-pruning}, fine-tuning~\cite{finetune}, and trigger filtering. We conduct experiments with the ResNet-18 model on the SPC-10 dataset for simplicity. Unless otherwise specified, all other settings are the same as those illustrated in Section \ref{experimentalsetting}.

\begin{figure*}[!t]
\begin{minipage}[t]{0.48\linewidth}{
\subfloat[PBSM]{
\includegraphics[width=0.473\columnwidth]{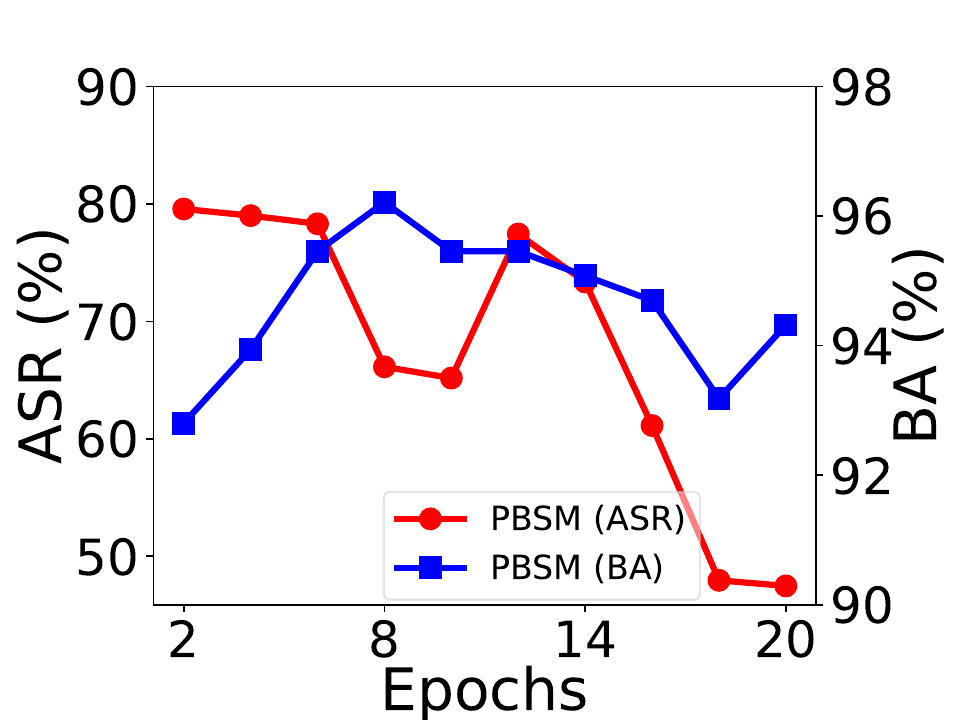}}
\subfloat[VSVC]{
\includegraphics[width=0.473\columnwidth]{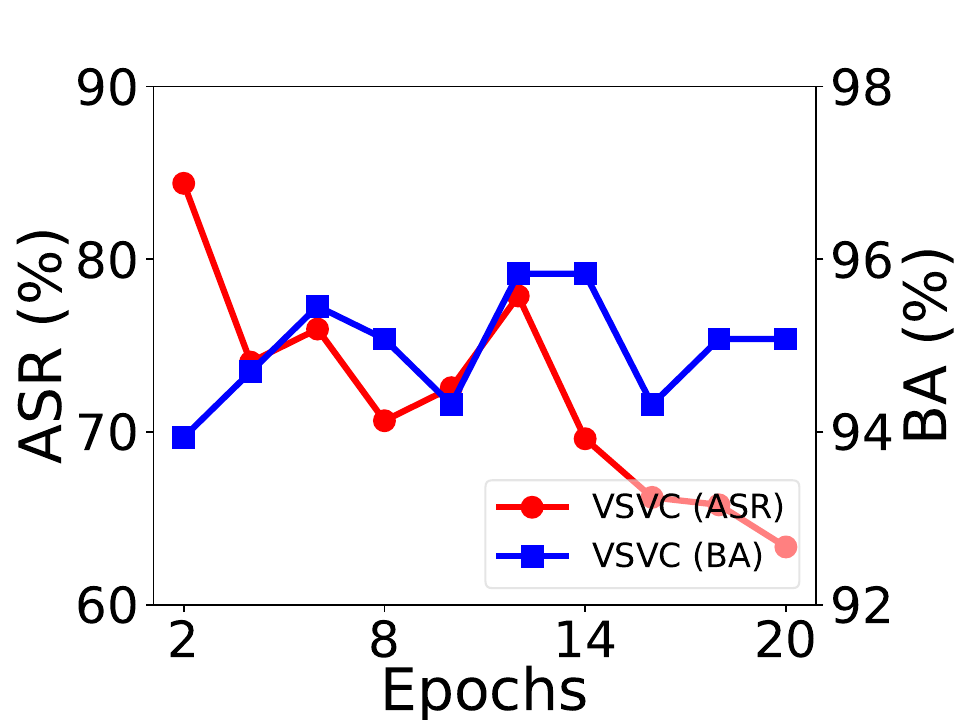}}
\caption{The resistance of our PBSM and VSVC to fine-tuning.}
\label{fig:Finetuning}}
\end{minipage}
\hfill
\begin{minipage}[t]{0.48\linewidth}{
\subfloat[PBSM]{
\includegraphics[width=0.473\columnwidth]{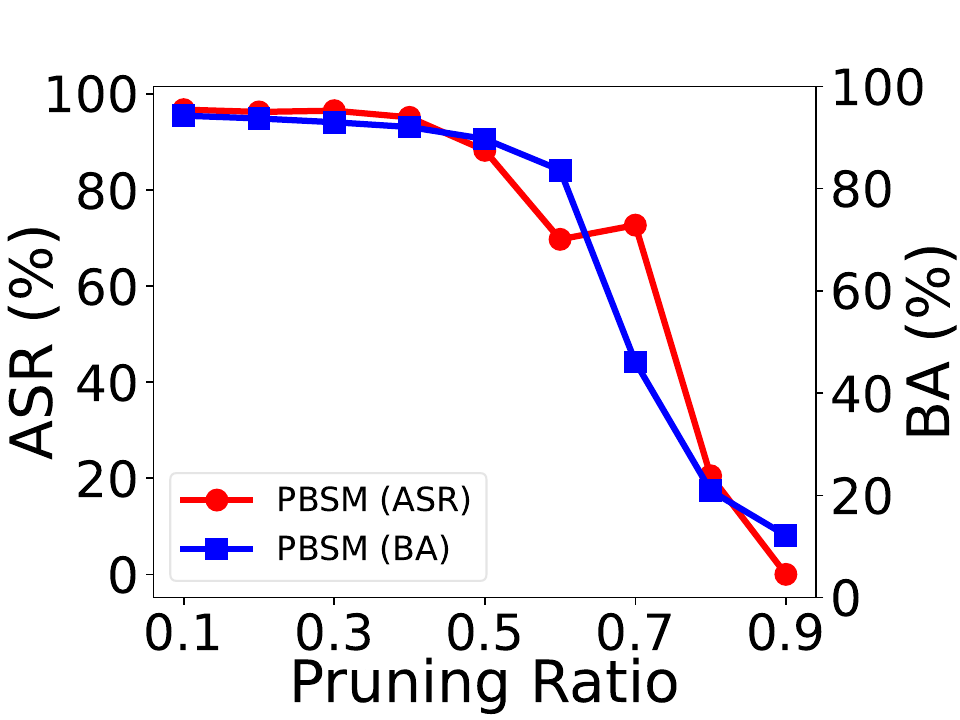}}
\subfloat[VSVC]{
\includegraphics[width=0.473\columnwidth]{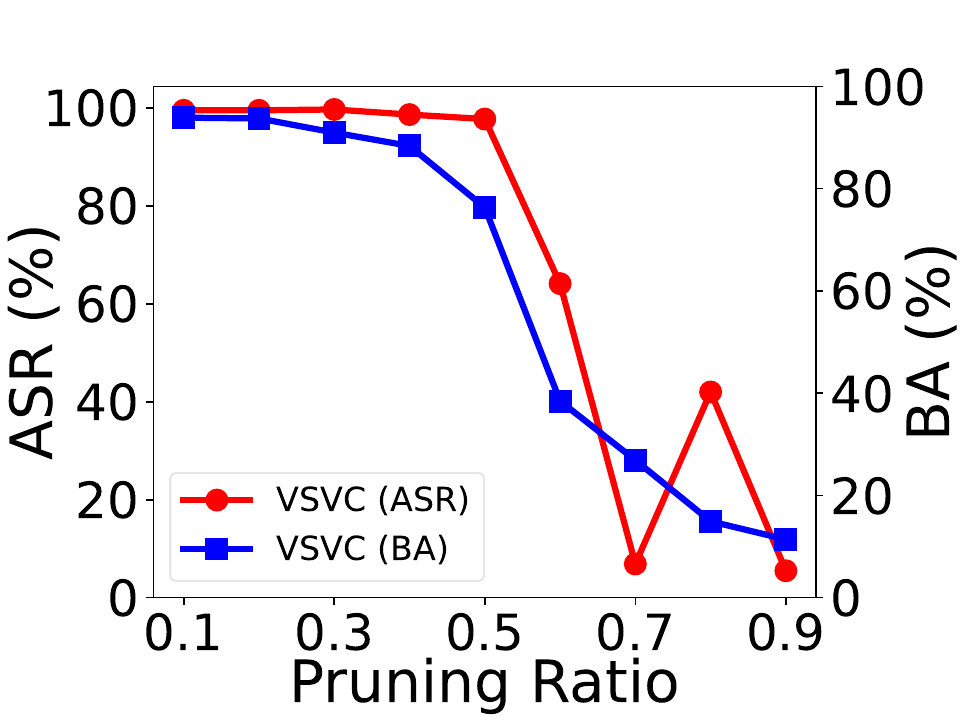}}
\caption{The resistance of our attacks to model pruning.}
\label{fig:Pruning}}
\end{minipage}
\end{figure*}

\vspace{0.3em}
\noindent \textbf{The Resistance to Fine-tuning.} 
As a representative backdoor-removal method, fine-tuning~\cite{finetune} intend to remove model backdoors by fine-tuning it with a few local benign samples. This method is motivated by the catastrophic forgetting property \cite{kirkpatrick2017overcoming} of DNNs. In our experiments, we exploit 10\% of benign training samples as our benign data and set the learning rate as 0.005. As shown in Figure~\ref{fig:Finetuning}, the attack success rate decreases with the increase of the tuning epoch. However, even at the end of this process, the ASRs are still larger than 45\% for both our PBSM and VSVC. These results verify that our attacks are resistant to fine-tuning to a large extent.

\vspace{0.3em}
\noindent \textbf{The Resistance to Model Pruning.} As another representative backdoor-removal defense, model pruning~\cite{fine-pruning} aims to remove model backdoors by pruning neurons that are dormant during the inference process of benign samples. This method is motivated by the assumption that backdoor and benign neurons are mostly separated in attacked DNNs. As shown in Figure~\ref{fig:Pruning}, the attack success rates are significantly decreased when pruning large amounts of neurons. However, it comes at the cost of a sharp decrease in benign accuracy. Specifically, the ASR decreases by almost the same amount as the BA for both PBSM and VSVC. This is mostly because the assumption of model pruning does not hold in our attacks due to their global and complex trigger designs. These results verify the resistance of our attacks to model pruning.

\begin{table}[!t]
\centering
\caption{The attack success rate (\%) of PBSM-infected DNNs on pitch-boosted samples with (w/) and without (w/o) injecting the high-pitch signal on SPC-10 and SPC-30 datasets.}
\vspace{-0.5em}
\label{tab:filter-defence}
\scalebox{1.2}{
\begin{tabular}{c|cc}
\toprule
\tabincell{}{Method$\rightarrow$\\Dataset$\downarrow$} & PBSM (w/o) & PBSM (w/) \\ \hline \rule{0pt}{8pt}
SPC-10  & 65.04\%   &  95.78\%                               \\ \hline \rule{0pt}{8pt}
SPC-30  & 70.62\%   &  96.63\%                               \\ 
\bottomrule
\end{tabular}}
\end{table}

\vspace{0.3em}
\noindent \textbf{The Resistance to Trigger-removal Defense.} To deactivate the potential backdoor in attacked DNNs, the defenders may remove its high-pitched signals, low-pitched signals, and noises, to remove potential trigger patterns of the suspicious testing audio. Obviously, this method has minor effects on our VSVC since we change the global features of its poisoned samples. However, it may defeat our PBSM since we inject a high-pitched signal after boosting the pitch. Accordingly, we examine whether our PBSM attack is still effective when using pitch-boosted samples without injecting the high-pitch signal to query the PBSM-infected DNNs. As shown in Table \ref{tab:filter-defence}, our attack can still reach satisfied attack success rates ($>65\%$) even without the high-pitch signals. It is mostly because our boosted pitch can also serve as a trigger pattern (as we mentioned in Section \ref{sec:PBSM}) which cannot be removed by trigger filtering. It verifies the resistance of our attacks again.

\subsection{Discussions}
In this section, we discuss the attack effectiveness of our methods under more difficult settings.

\begin{figure}[!t]
\begin{minipage}[t]{0.48\linewidth}
\centering
\includegraphics[width=\columnwidth]{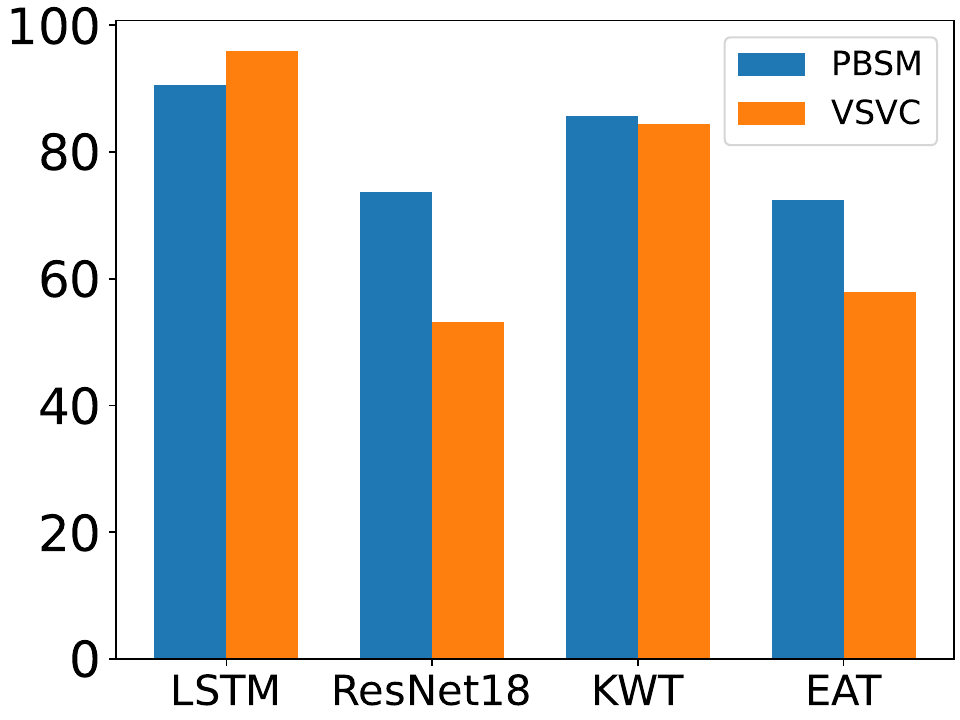}
\caption{Clean-Label Attacks.}
\label{fig:clean_label}
\end{minipage}%
\hfill
\begin{minipage}[t]{0.48\linewidth}
\centering
\includegraphics[width=\columnwidth]{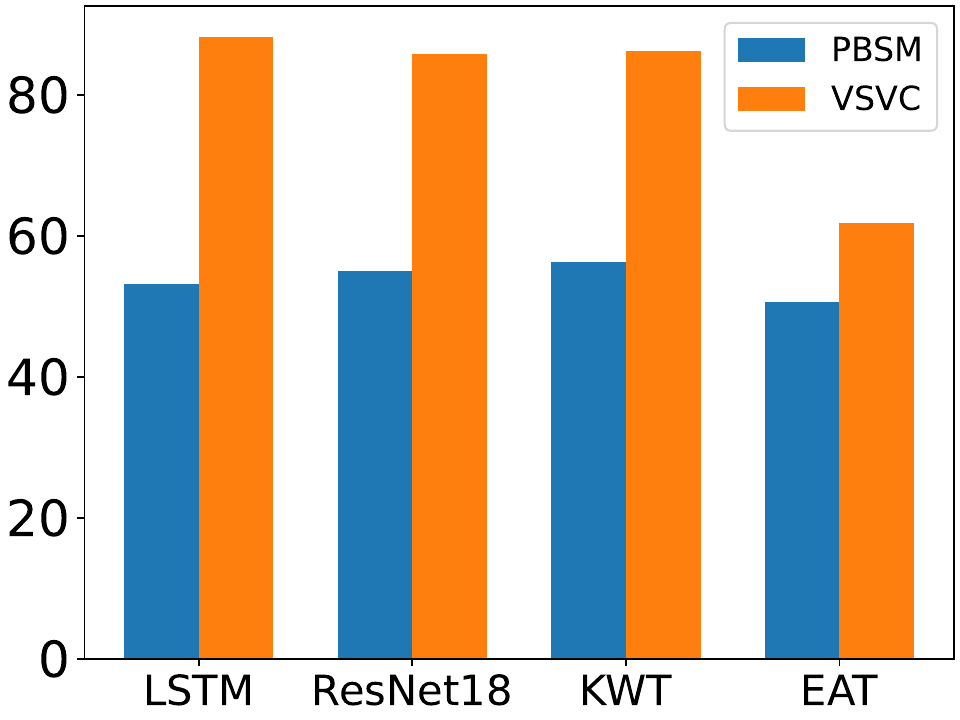}
\caption{Over-the-Air Attacks.}
\label{fig:over_air}
\end{minipage}
%\vspace{-5mm}
\end{figure}

\vspace{0.3em}
\noindent \textbf{Attacks under the Clean-Label Setting.}
Although our attacks are imperceptible, the label of the poisoned samples usually differs from that of their clean versions. Accordingly, users may identify the attack by inspecting the audio-label relation when they can catch some poisoned samples. To further demonstrate the effectiveness of our methods, we explore whether they are still effective under the clean-label setting. In these experiments, we only select samples from the target class for poisoning instead of sampling data from all classes and changing their label to the target one. As shown in Figure \ref{fig:clean_label}, although the performances are relatively weaker than those of attacks under the poisoned-label setting, our attacks are still effective when poisoning 9\% samples. Specifically, the average ASRs across all model structures of PBSM and VSVC are 81\% and 73\%, respectively. These results verify the effectiveness of our PBSM and VSVC under the clean-label setting.

\vspace{0.3em}
\noindent \textbf{Attacks under the Over-the-Air (Physical) Setting.} 
To evaluate the effectiveness of our attack methods in real-world scenarios, we design a physical experiment to assess the performance of our attacks under the over-the-air setting. Specifically, we conduct these experiments in a room, where we use computer speakers to play our backdoor audio and a smartphone is used as the recording device to capture the audio. The obtained audio is input into the attacked DNNs for prediction. We measure the playback volume of the audio and it is similar to that of a normal conversation. We place the smartphone at a distance of 0.5 meters from the speaker. As shown in Figure~\ref{fig:over_air}, although the performances are relatively weaker than those of attacks under the digital setting, our attacks are still effective in the real world. Specifically, the average ASRs across all model structures of PBSM and VSVC are 53\% and 80\%, respectively. The lower ASR of the PBSM is mostly due to the limitations of our evaluated device, which may not effectively capture high-pitched signals.

% \usepackage{multirow}
\begin{table}[!t]
\centering
\caption{The performance of PBSM and VSVC under the all-to-all setting on SPC-10. In this table, we provide the result per class following the settings in BadNets \cite{gu2019badnets}. }
\vspace{-0.5em}
\label{tab:all-to-all}
\scalebox{0.97}{
\begin{tabular}{cc|cc|cc}
\toprule
\multirow{2}{*}{Class} & \multirow{2}{*}{\begin{tabular}[c]{@{}c@{}}Accuracy (\%) of \\ Benign Model \\ \end{tabular}} & \multicolumn{2}{c|}{PBSM} & \multicolumn{2}{c}{VSVC} \\
                       &                                                                                  & BA (\%)         & ASR (\%)       & BA (\%)         & ASR (\%)       \\ \hline \rule{0pt}{8pt}
yes                    & 95.70                                                                            & 95.71       & 92.19      & 93.75       & 93.36      \\
left                   & 97.38                                                                            & 96.26       & 91.39      & 96.63       & 95.13      \\
off                    & 96.57                                                                            & 95.04       & 90.08      & 96.18       & 89.70      \\
on                     & 97.97                                                                            & 97.15       & 96.34      & 96.75       & 94.72      \\
go                     & 94.82                                                                            & 94.42       & 84.06      & 90.44       & 81.28      \\
down                   & 92.10                                                                            & 93.28       & 89.33      & 89.72       & 90.12      \\
stop                   & 96.79                                                                            & 97.59       & 93.17      & 95.58       & 94.38      \\
no                     & 90.87                                                                            & 90.48       & 84.52      & 90.87       & 81.35      \\
right                  & 96.53                                                                            & 95.75       & 92.66      & 94.21       & 93.45      \\
up                     & 97.11                                                                            & 98.16       & 93.02      & 96.32       & 92.65      \\  \bottomrule
\end{tabular}}
\end{table}

\vspace{0.3em}
\noindent\textbf{Attacks under the All-to-All Setting.} 
To further illustrate the effectiveness of our PBSM and VSVC, we extend the all-to-one attack setting to a more challenging all-to-all one, where the target label $y_t$ of a poisoned sample (with ground-truth class $y$) is set to $y' = (y+1) \mod K$. In particular, we increase the poisoning rate to 15\% due to the difficult of this task. We conduct experiments on the SPC-10 dataset with ResNet-18. As shown in the Table~\ref{tab:all-to-all}, both PBSM and VSVC can reach promising performance against samples from all classes, although the performance may have some mild fluctuations across them. These results confirm the feasibility of our attacks under the all-to-all setting.

\subsection{Analyzing Attacks in the Hidden Feature Space}
In this section, we analyze why our PBSM and VSVC attacks are effective from the behaviors of samples in the hidden feature space of attacked DNNs.

\vspace{0.3em}
\noindent \textbf{Settings.} In this section, we visualize the features of poisoned samples generated by the backbone (\ie, the input of fully-connected layers) of attacked DNNs via t-SNE~\cite{t-SNE}. For simplicity, we adopt 2,500 samples and exploit ResNet-18 trained on the SPC-10 dataset for our analysis. 

\begin{figure}[!t]
\centering
\subfloat[PBSM]{\label{fig:subfig:PBSM_TSNE}
\includegraphics[width=0.45\columnwidth]{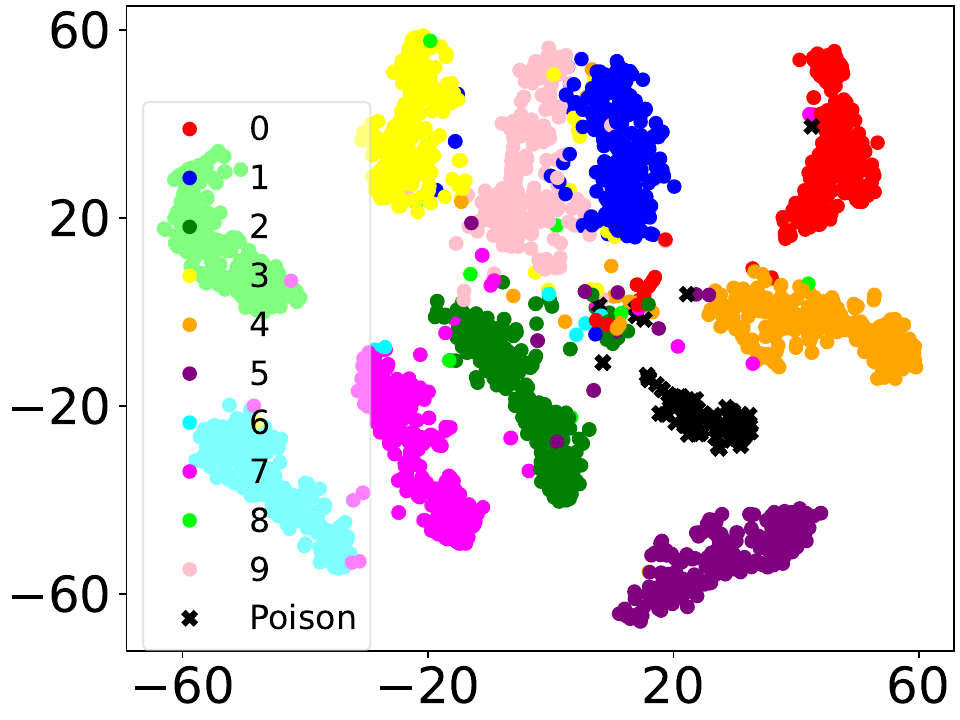}}\hspace{0.4em}
\subfloat[VSVC]{\label{fig:subfig:VSVC_TSNE}
\includegraphics[width=0.45\columnwidth]{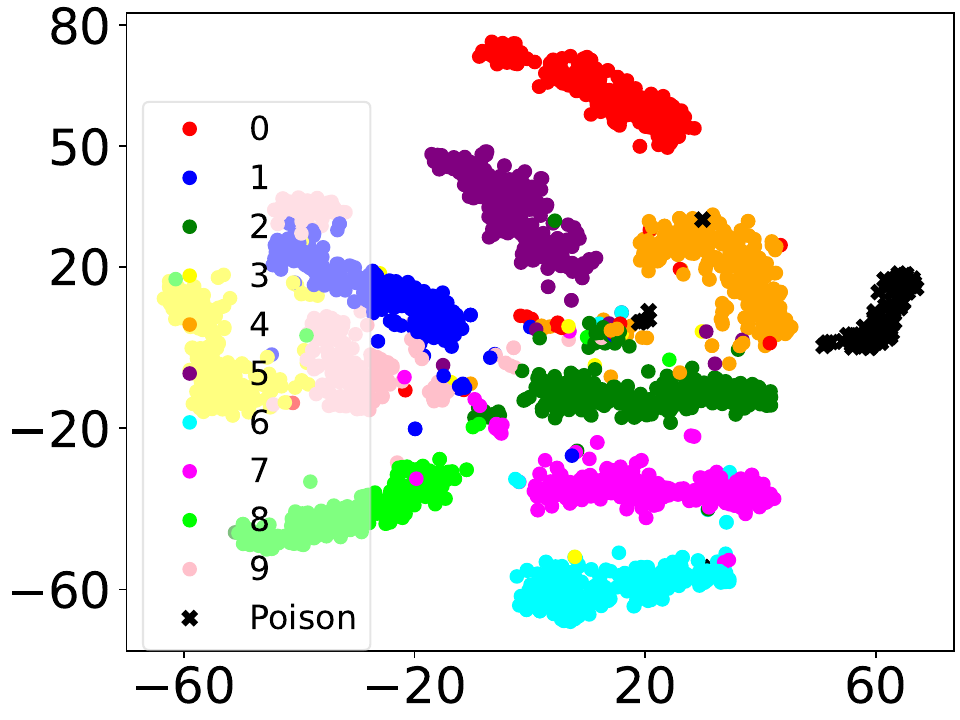}}
\caption{The t-SNE visualization of features of benign and poisoned samples from the hidden feature space generated by PBSM-infected and VSVC-infected models.}\label{fig:TSNE}
\end{figure}

\vspace{0.3em}
\noindent \textbf{Results.} As shown in Figure~\ref{fig:TSNE}, poisoned samples (marked in black) cluster together regardless of their ground-truth labels. In contrast, the benign samples form separate clusters according to their ground-truth class. These phenomena are consistent with predicted behaviors of the attacked model where it `assigns' the same label to all samples in the same cluster. These results also verify the effectiveness of our attacks, showing that they can force attacked DNNs to learn features of triggers and ignore the benign features. It enables attacked DNNs to minimize the distance between poisoned samples in the feature space and associate the learned trigger-related features with the target label.

\section{Conclusion}
\label{sec:conclusion}
In this paper, we revealed that almost all existing poison-only backdoor attacks against speech recognition are not stealthy due to their simple trigger designs. To overcome this deficiency, we proposed two simple yet effective attacks, including pitch boosting and sound masking (PBSM) and voiceprint selection and voice conversion (VSVC), inspired by the elements of sound. Our attacks generated more `natural' poisoned samples and therefore are more stealthy. We also generalized and evaluated our attacks under more difficult settings, such as all-to-all, clean-label, and physical ones. However, we notice that the attack performance may have some degrades in some cases under these settings. We will explore how to alleviate this problem and design their defense countermeasures in our future works. We hope that our research can provide a deeper understanding of stealthy backdoor attacks in speech recognition, to facilitate the design of more sure and robust speech recognition models.

\bibliographystyle{IEEEtran}
\bibliography{mybib}

\begin{IEEEbiography}[{\includegraphics[width=1in,height=1.25in,clip,keepaspectratio]{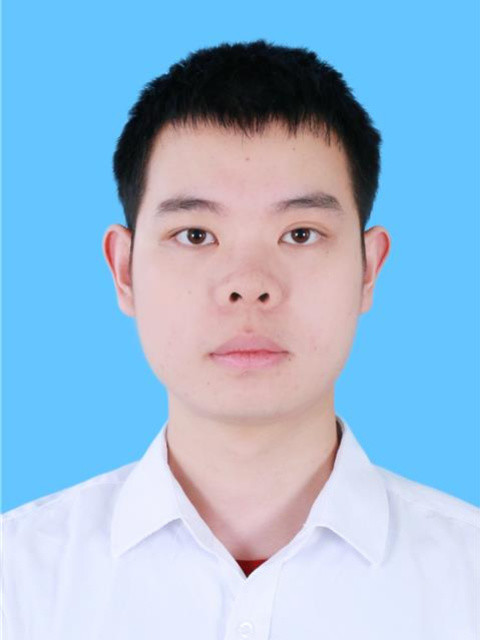}}]{Hanbo Cai} is currently a Ph.D. candidate in Computer Science and Technology, College of Computer and Information, Hohai University. His research interests primarily focus on AI security, particularly backdoor learning and adversarial attacks.

\end{IEEEbiography}

\begin{IEEEbiography}
[{\includegraphics[width=1in,height=1.25in,clip,keepaspectratio]{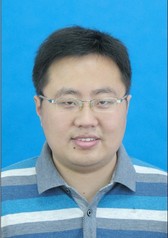}}]{Dr. Pengcheng Zhang} received the Ph.D. degree in computer science from Southeast University in 2010. He is currently a full professor in College of Computer and Information, Hohai University, Nanjing, China. His research interests include software engineering, service computing and data science. He has published research papers in premiere or famous computer science journals, such as IEEE TSE, IEEE TSC, IEEE TKDE, IEEE TBD, IEEE TETC, IEEE TCC and IEEE TR. He was the co-chair of
IEEE AI Testing 2019 conference. He served as a technical program committee member on various international conferences. %He is a member of the IEEE.
\end{IEEEbiography}

\begin{IEEEbiography}
[{\includegraphics[width=1in,height=1.25in,clip,keepaspectratio]{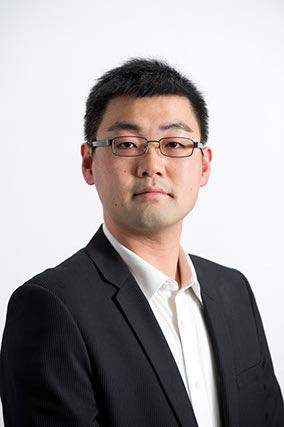}}]{Dr. Hai Dong} received a PhD from Curtin University, Perth, Australia. He is currently a senior lecturer at School of Computing Technologies in RMIT University, Melbourne, Australia. He was previously a Vice-Chancellor's Research Fellow in RMIT University and a 
Curtin Research Fellow in Curtin University. His primary research interests include: Services Computing, Edge Computing, Blockchain, Cyber Security, Machine Learning and Data Science. His publications appear in ACM Computing Surveys,  IEEE TIE, IEEE TII, IEEE TSC, IEEE TSE, etc. He is a Senior Member of the IEEE.
\end{IEEEbiography}

\begin{IEEEbiography}
[{\includegraphics[width=1in,height=1.25in,clip,keepaspectratio]{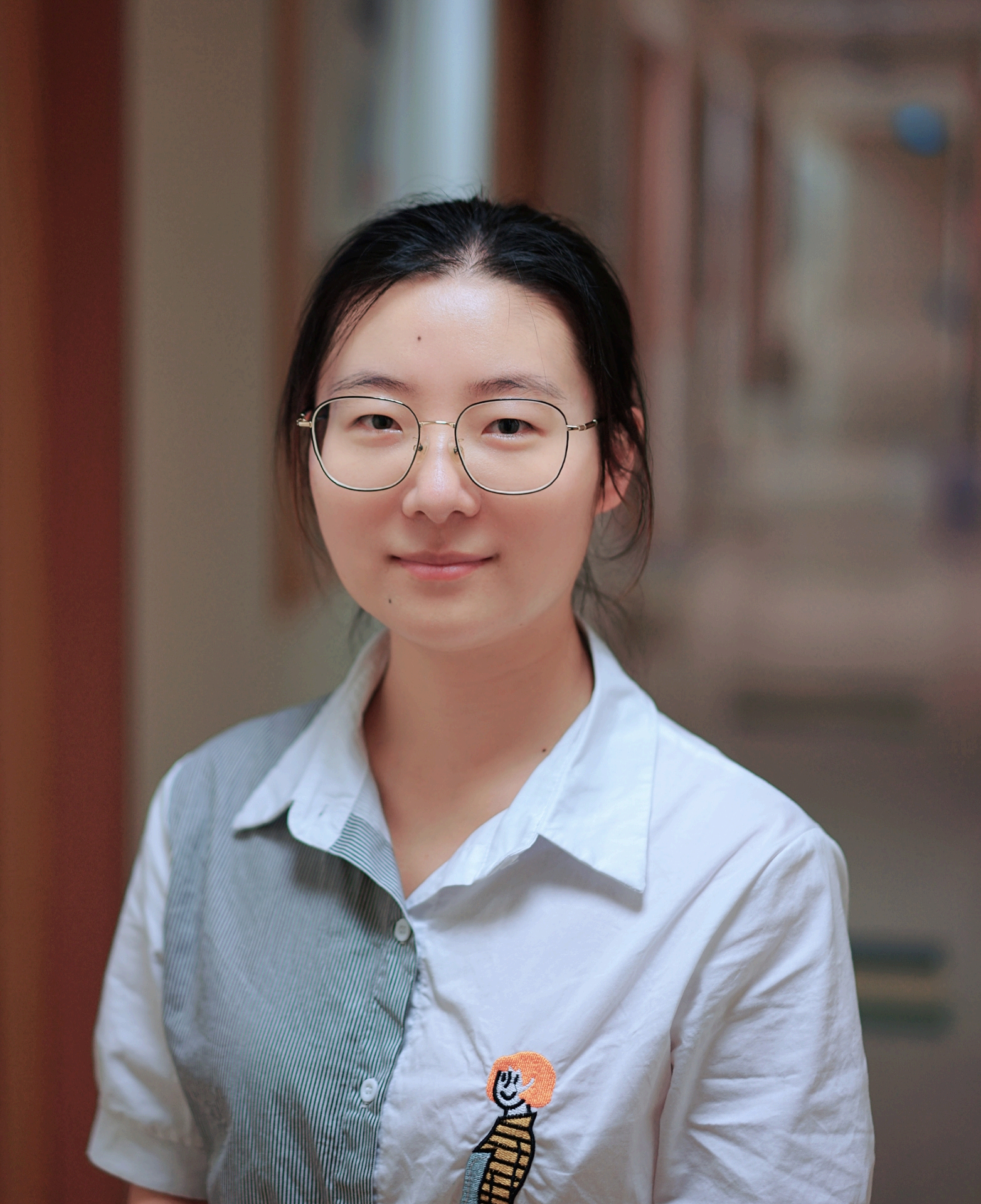}}]{Dr. Yan Xiao} is an Associate Professor at School of Cyber Science and Technology in Sun Yat-sen University. She received her PhD degree from the City University of Hong Kong and held a research fellow position at National University of Singapore. Her research focuses on trustworthiness of deep learning systems and AI applications in software engineering. More information is available on her homepage: https://yanxiao6.github.io/.
\end{IEEEbiography}

\begin{IEEEbiography}
[{\includegraphics[width=1in,height=1.25in,clip,keepaspectratio]{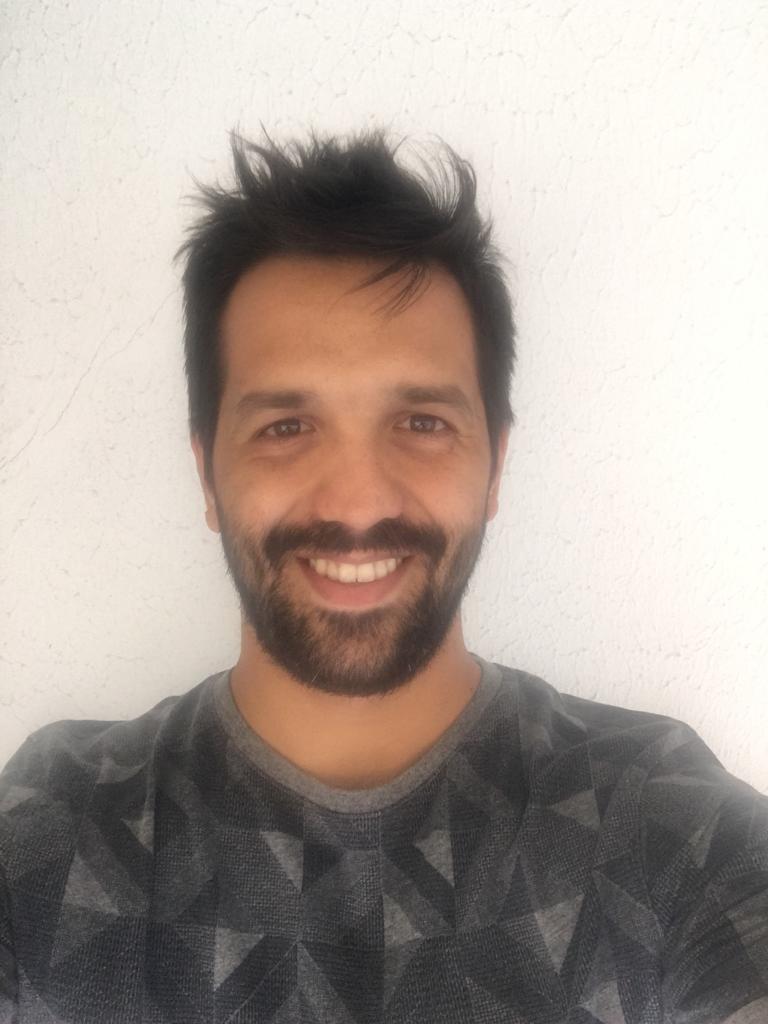}}]{Stefanos Koffas} is currently a Ph.D. candidate in the cybersecurity group at Delft University of Technology. His research focuses on the security of AI and especially on backdoor attacks in neural networks. Before that, he obtained his MSc. in Computer Engineering from Delft University of Technology and his M.Eng. in electrical and computer engineering from National Technical University of Athens, Greece.
\end{IEEEbiography}

\begin{IEEEbiography}
[{\includegraphics[width=1in,height=1.25in,clip,keepaspectratio]{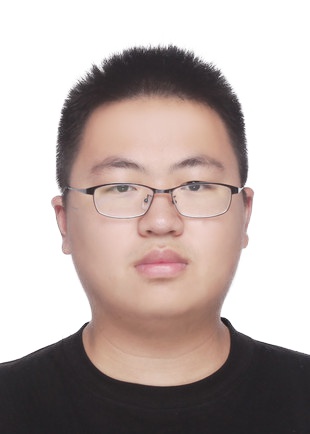}}]{Dr. Yiming Li} received his Ph.D. degree in Computer Science and Technology from Tsinghua University in 2023. Before that, he received his B.S. degree in Mathematics and Applied Mathematics from Ningbo University in 2018. His research interests are in the domain of Trustworthy ML, especially backdoor learning and copyright protection in deep learning. His research has been published in multiple top-tier conferences and journals, such as ICLR, NeurIPS, ICCV, ECCV, IEEE TIFS and IEEE TNNLS. He served as the senior program committee member of AAAI, the program committee member of ICLR, NeurIPS, ICML, etc., and the reviewer of IEEE TPAMI, IEEE TIFS, IEEE TDSC, etc. His research has been featured by major media outlets, such as IEEE Spectrum. He was the recipient of the Best Paper Award in PAKDD 2023.
\end{IEEEbiography}

\vfill

\end{document}